# Static Polarizabilities of Dielectric Nanoclusters

Hye-Young Kim[a,c] , Jorge O. Sofo[a,c] , Darrell Velegol[b,c] , Milton W. Cole[a,c]

*Department of Physics[a] , Chemical Engineering[b] and the Materials Research Institute[c]*
*The Pennsylvania State University*
*University Park, PA 16802*

and Gautam Mukhopadhyay
*Physics Department*
*Indian Institute of Technology-Bombay*
*Powai, Mumbai 400076 India*

Submitted date: July 8, 2005

ABSTRACT

A cluster consisting of many atoms or molecules may be considered, in some circumstances, to be a single large molecule with a well defined polarizability. Once the polarizability of such a cluster is known, one can evaluate certain properties, e.g. the cluster's van der Waals interactions, using expressions derived for molecules. In the present work, we evaluate the static polarizability of a cluster using a microscopic method that is exact within the linear and dipolar approximations. Numerical examples are presented for various shapes and sizes of clusters composed of identical atoms, where the term "atom" actually refers to a generic constituent, which could be any polarizable entity. The results for the clusters' polarizabilities are compared with those obtained by assuming simple additivity of the constituents' atomic polarizabilities; in many cases, the difference is large, demonstrating the inadequacy of the additivity approximation. Comparison is made (for symmetrical geometries) with results obtained from continuum models of the polarizability. Also, the surface effects due to the nonuniform local field near a surface or edge are shown to be significant.

## I.    INTRODUCTION

The polarizability is one of the important properties of atoms and molecules, with a wide variety of applications [1, 2]. For example, we have used the atomic or molecular polarizability in computations of the van der Waals (VDW) interaction between clusters of various shapes by adding two-body and three-body VDW interactions between constituent atoms [3-5].  A more efficient way (exact at large separation) to evaluate the VDW interaction between clusters of arbitrary shape would be to consider each cluster as a single large molecule with a well defined polarizability tensor and use expressions derived for the VDW interaction between molecules. There is a body of experimental and theoretical information concerning these quantities, primarily for the case of metallic clusters [6-8]. The present study was motivated by the desire to extend that database to include dielectric clusters. We have carried out that extension by a series of calculations based on either continuum models or a microscopic theory, describing discrete models of atoms located at a set of lattice sites. By the term "atom" we mean an entity lacking a permanent dipole moment; the case of such moments requires separate treatment, which will be undertaken in the future.

In the next section, we present the basic formalism of the microscopic method and evaluate the static polarizabilities of clusters composed of a single species of atom. For a given cluster, this is achieved by determining self-consistently the induced dipole moment of each constituent atom, which is polarized by the local electric field. The latter is a sum of the applied external electric field and the internal field provided by induced dipole moments of neighboring atoms [2, 9-11]. This method is exact in the linear and dipolar approximation limit. An alternative continuum model approach yields values that can be compared with the discrete model. The main results are (i) numerical results for the dependence of the cluster polarizability tensor on size, shape, orientation, and substance, (ii) analytical results for the asymptotic polarizability in the large size limit, and (iii) a description of the surface effect due to the nonuniform local field within the cluster. Our results are summarized and diverse applications are discussed in Section III.

## II.    THE STATIC POLARIZABILITY OF A ONE COMPONENT CLUSTER

The idea behind the microscopic method we employ is simple [2, 9-11]. When a static external electric field is applied to a cluster of atoms, each atom will develop an induced dipole moment that will depend on both the applied field and the field resulting from all other induced dipoles. This local-field effect will be responsible for all particles not having an identical polarization. The result of taking mutual interactions into account is a different polarizabilty for the cluster from that one would obtain from simple addition of atomic polarizabilities. The formalism of this microscopic method is straightforward and briefly summarized in order to introduce the notation.

The local microscopic electric field ($\mathbf{E}^{loc}$) at the site of an atom within a cluster is generally different from the macroscopic electric field. In general, the local field at the position of the j-*th* atom ($\mathbf{E}_j^{loc}$) is a sum of the applied macroscopic electric field ($\mathbf{E}_0$), originating from an external source, and the internal field at that site ($\mathbf{E}_{\text{int},j}$) due to the field created by the induced dipoles of neighboring atoms [9-11]:

$$\mathbf{E}_j^{loc} = \mathbf{E}_0 + \mathbf{E}_{\text{int},j} \tag{1}$$

$$\mathbf{E}_{\text{int},j} = \sum_{\substack{i=1 \\ i \neq j}}^{N} \frac{(3\mathbf{n}_{ji}\mathbf{n}_{ji} - \mathbf{I}) \cdot \mathbf{p}_i}{r_{ji}^3} \equiv \mathbf{T}_{ji} \cdot \mathbf{p}_i \tag{2}$$

Here, $r_{ji} = \left| \mathbf{r}_j - \mathbf{r}_i \right|$, $\mathbf{T}_{ji}$ is the dipole field tensor, the unit separation vector $\mathbf{n}_{ji} = (\mathbf{r}_j - \mathbf{r}_i)/r_{ji}$, and $\mathbf{p}_i$ is the induced dipole moment of the i-*th* atom in the cluster of $N$ atoms. The sum is over all neighboring atoms *i* within the cluster and the usual summation convention over repeated identical indices is employed. The induced dipole moment of the cluster ($\mathbf{p}_{cluster}$) is the sum of the induced dipole moments of the constituent atoms:

$$\mathbf{p}_{cluster} = \sum_{j=1}^{N} \mathbf{p}_j \quad . \tag{3}$$

In general, for a cluster consisting of polarizable atoms the polarizability tensor of the cluster ($\boldsymbol{\alpha}_{cluster}$) is defined as the coefficient of proportionality relating the induced dipole moment of the cluster to the applied external field:

$$\mathbf{p}_{cluster} = \boldsymbol{\alpha}_{cluster} \cdot \mathbf{E}_0 \quad . \tag{4}$$

Hence, one may obtain the polarizability from Eqs. (3) and (4) by evaluating the induced moment on each atom in the cluster. The moment on the j-*th* atom is

$$\mathbf{p}_j = \alpha_{atom} \mathbf{E}_j^{loc} \tag{5}$$

where $\alpha_{atom}$ is the atomic polarizability. Since each dipole reacts to induced moments on the neighboring atoms, the moment of each atom must be calculated self-consistently. By letting $\mathbf{p}_j{'} \equiv \dfrac{\mathbf{p}_j}{\alpha_{atom} E_0}$ and $\mathbf{E}_0 = E_0 \mathbf{e}$, one obtains *3N* linear equations from Eqs. (1), (2) and (5):

$$\mathbf{p}_j{'} - \alpha_{atom} \mathbf{T}_{ji} \cdot \mathbf{p}_i{'} = \mathbf{e} \quad . \tag{6}$$

The numerical solution of these coupled equations yields the induced dipole moment of each atom in the cluster [2]. One can obtain all the nine elements of the polarizability tensor $\boldsymbol{\alpha}_{cluster}$ by rotating the cluster with respect to the direction of the external applied field. For completeness, we summarize the key approximations used to obtain this polarizability: static linear response, the assumption of point dipoles on lattice sites, the neglect of higher order multipoles, and the assumption that the atomic polarizabilities are not affected by their environments.

For comparison with this result for $\boldsymbol{\alpha}_{cluster}$, we evaluate a naive approximation to the cluster polarizability by ignoring local fields and adding the individual atomic polarizabilities. Within this "additivity approximation", the polarizability $\boldsymbol{\alpha}_{cluster}$ would equal $N\alpha_{atom}$. Therefore, in the following, we focus our attention on a dimensionless ratio *f*, the "enhancement factor":

$$\mathbf{f} \equiv \frac{\alpha_{cluster}}{N\alpha_{atom}} \qquad .$$
(7)

In spite of the name, we will find values of $f > 1$ for some situations while $f < 1$ for others. One might refer to the latter case as "screening" and the former case as "anti-screening".

In the present work, the static polarizabilities of clusters are investigated for several shapes, sizes and orientations: linear clusters (with dimensions 1x1xL) over the range $2 \leq L \leq 4000$, square prisms (2x2xL) having $2 \leq L \leq 250$, cubic clusters (LxLxL) with $2 \leq L \leq 17$, and square monolayer clusters (1xLxL) with $2 \leq L \leq 70$. The static polarizabilities of spherical clusters with total number of atoms $N \leq 5000$ are also calculated, for which an exact analytic solution is available from continuum theory. Here, each cluster consists of atoms residing at simple cubic lattice sites with lattice constant $a_0$ (see Table I), the value of which is obtained from the known bulk density $n_s$ of each substance[5]; the lattice constant is the unit of length, except for spheres. For a better description of the spherical configuration, face centered cubic lattice sites are adopted for spherical clusters, keeping the same bulk density. The calculations proceeded over the specified range of size up to saturation at an asymptotic limit, or else (in the case of square monolayer clusters) until a computational limit arose.

For quantitative estimates, we use atomic polarizabilities and lattice constants of substances such as silica, hexane and sapphire, as listed in Table I [5]. These "atomic" polarizabilities are derived using the Clausius-Mossotti (CM) relation and known dielectric spectra for n-hexane ($C_6H_{14}$) [12], fused silica ($SiO_2$) [12] and sapphire (>99.9% $Al_2O_3$)[12]. The "atomic" polarizabilities are taken from the entire unit (e.g., $SiO_2$, $Al_2O_3$). The dielectric spectra come from absorption measurements, giving the loss modulus ($\varepsilon''$) at real frequencies ($\omega$); a Kramers-Kronig relation then transforms this function to the real function at imaginary frequency, $\varepsilon(i\omega)$. The complexity of the spectra makes it inappropriate to use an alternative, simple Drude model.

In Fig. 1, values of the computed enhancement factor $f$ for cubic clusters are shown as a function of the total number of atoms in the cluster $N$. Due to symmetry, the polarizability of a cubic cluster is isotropic. As the cluster size increases, $f$ increases and approaches an asymptote that depends on the specific substance. For clusters with asymmetric shapes, $\mathbf{\alpha}_{cluster}$ is anisotropic and $f$ has a striking orientation dependence to its variation with $N$ (which will be explained below). For either linear clusters or square prisms, $f$ increases (decreases) with $N$ as the long dimension of the cluster lies along (perpendicular to) the external applied field, approaching an asymptotic value, as shown in Figs. 2 and 3. Qualitatively similar anisotropic behavior is also observed for a square monolayer cluster, as shown in Fig. 4. There is a consistent and expected dependence on the substance polarizability for all clusters with various shapes, sizes and orientations. Specifically, these exhibit successively larger values of $\left| f - 1 \right|$ in the order hexane, silica and sapphire. This ordering is that of the product $n_s \alpha_{atom}$ (as listed in Table I), indicating that large magnitude corrections to additivity occur when the atomic polarizability per unit volume of the cluster is large, as expected.

In addition to the approach to the asymptotic limits, one other general observation can be made about the data in Figs. 1-4. For $f>1$ cases, there is an inflection point for the curve at N $\leq$ 10 for all shapes and substances, while for $f<1$ cases there is no such inflection point observed. (An exception is the sapphire disc cluster lying along the external field and the reason for this is addressed below.) This presence or absence of an inflection point may be explained as a result of the nonuniform local field at the surface of the cluster. One of the outputs from our calculation is the magnitude and orientation of the induced dipole moment on each atomic site. The distribution of the induced dipole moments within a cluster is found to be nonuniform and clearly shows this surface effect. In Figs. 5 and 6, the distributions of magnitude of induced dipole moments are shown for a silica square monolayer cluster ($L$=60) lying perpendicular to and along the external field, respectively. One observes the "local" enhancement factor (defined as $f_j^{'} \equiv \dfrac{\mathbf{p}_j}{\alpha_{atom} E_0}$ ) instead of the actual magnitude of induced dipole moments ( $\mathbf{p}_j$ ) at each site $j$.

When the square monolayer lies perpendicular to the external field, all of the induced moments are parallel to the same direction as the applied external field but the magnitude varies with position. On the other hand, when the field lies in the plane of the square monolayer, there is an additional nonuniform orientation (Fig. 7a), as indicated schematically with arrows in Fig. 7(b) (reminiscent of the field lines of a giant dipole). For a better quantitative understanding, Figs. 8-10 show the magnitude and orientation of induced moments for silica square monolayer of various sizes in two orientations. Similarly, Figs. 11 and 12 show those for silica linear clusters of various lengths in two orientations. Fig. 12(b) shows behavior for the inner region close to the edges as function of the distance from the edge in unit of lattice spacing for L=50 to 4000 where all the data points overlap. It is clear that there is a surface layer, where the moment varies rapidly, before reaching the uniform bulk value. We will call this thickness the "penetration depth". Note that this depth is very small (~ 1 $a_0$) when the field is parallel to the given edge (Figs. 8, 9(a) and 11) . This depth is much greater when the edge is perpendicular to the field. The value is ~ 6 $a_0$ for linear clusters (Fig. 12) and ~ 10 $a_0$ for square monolayers (Fig. 9(b)). The different behavior for the two edge orientations may be explained by comparing the coupling between two simple pairs of identical dipoles ( **p** ) separated at a distance *d*; one pair lying along and another pair lying perpendicular to the orientation of dipole moments. When the dipoles lie along (perpendicular to) the direction of their dipole moments, the interaction is attractive (repulsive) and the magnitude is *-2p²/d³* ( *p²/d³* ). These two pairs correspond to the induced dipoles in a cluster aligned along and perpendicular to the external field, respectively. Therefore, the dipole interaction due to neighboring dipoles in the former case is stronger (by factor of 2) and thus effectively extends over longer range, resulting in a smooth conversion over the larger penetration depth than that in the latter case. Secondly, the penetration depth of 6 and 10 atomic lattice spacing for linear and square monolayer clusters in case of *f*>1 corresponds well to the size of the cluster where the inflection point is observed in Figs.1-4. On the other hand, in the orientation with *f*<1 where no inflection point is observed, the penetration depth is found to be only one atomic lattice spacing. Therefore, we reach a plausible conclusion that the inflection points indicate the critical size where the penetration depth becomes the same order of magnitude as the cluster size.

For a comparison, the polarizability of a spherical cluster is calculated. Unlike the cubic cluster, for large *N*, *f* decreases as *N* increases [see Fig. 13(a)]. The data in Fig. 13 show some "scatter" because the discrete lattice representation of the sphere is oversimplified for small radius. Note that, in Fig. 13(b), the size dependence of | *f*-1| is shown to be proportional to $R^{-1}$, which is the ratio of the surface area to the volume of the sphere while the magnitude *f* stays very close to the analytic result of 1 (as derived from continuum theory below). This empirical trend demonstrates that the deviation of *f* from unity arises from the surface penetration region.

We have investigated the asymptotic values of *f* by evaluating the polarizability in two other ways: one is an analytic solution for an infinite size cluster (i.e., $L = \infty$ ) in which each atom is assumed to have the same dipole moment, ignoring the surface effect. The other basis for comparison is an analytic solution, based on continuum theory, where a cluster is taken to be a dielectric ellipsoid [13]. The derivations of the analytic expressions listed in Table II are presented below.

The polarizability of an infinitely long chain of polarizable atoms lying along the external applied field is evaluated as follows. The moment on each atom is assumed to point along the chain; because each moment has the same magnitude, the result is evaluated from Eq. (6),

$$\mathbf{p}_{atom} = \alpha_{atom}\left[\mathbf{E}_0 + \frac{2\mathbf{p}_{atom}}{a_0^3}\left(\sum_{\substack{j=-\infty \\ j\neq 0}}^{\infty}\frac{1}{j^3}\right)\right] \qquad (8)$$

Here, the summation is over all neighboring sites, so it equals $2\zeta(3)$ where $\zeta(3) \approx 1.20205$ is the Riemann zeta function[14]. Now, the dipole moment on an atom becomes

$$\mathbf{p}_{atom} = \frac{\alpha_{atom}\mathbf{E}_0}{1 - \frac{4\alpha_{atom}}{a_0^3}\zeta(3)} \qquad . \qquad (9)$$

Hence, the polarizability of the infinite chain satisfies

$$f_{line}^{//E}(1\times1\times\infty) \equiv \frac{\alpha_{cluster}}{N\alpha_{atom}} = \frac{1}{1 - 4\zeta(3)n_s\alpha_{atom}}, \qquad (10)$$

where, $n_s = a_0^{-3}$ is used. The polarizabilities of other infinite size clusters are evaluated similarly and listed in Table II.

The polarizability along one of the principal axes of a dielectric ellipsoid (with semiaxes $a \geq b \geq c$) when an external field is applied parallel to one of its principal axes is given by [15]:

$$\alpha_1 = \frac{abc}{3}\frac{\varepsilon-1}{1+L_1(\varepsilon-1)}$$

$$\alpha_2 = \frac{abc}{3}\frac{\varepsilon-1}{1+L_2(\varepsilon-1)} \qquad (11)$$

$$\alpha_3 = \frac{abc}{3}\frac{\varepsilon-1}{1+L_3(\varepsilon-1)}$$

where the geometrical factors are $L_1 = Q(a,b,c)$, $L_2 = Q(b,c,a)$ and $L_3 = Q(c,a,b)$. Here

$$Q(a,b,c) = \frac{bc}{2a^2} \int\limits_0^\infty dx \; G^{-1/2}(x) \qquad\qquad, \tag{12}$$

$$G(x) = (x+1)^3 \left( x + (b/a)^2 \right) \left( x + (c/a)^2 \right) \tag{13}$$

and

$$L_1 + L_2 + L_3 = 1 \qquad . \tag{14}$$

Here, $\varepsilon$ is the dielectric constant of the substance. The geometrical factors for a sphere (a=b=c) are $L_1 = L_2 = L_3 = \frac{1}{3}$. An ellipsoid with shape similar to a linear cluster or a square prism is a prolate spheroid (i.e., a needle with b=c and a>> b, c) and an ellipsoid that is similar to a square monolayer cluster is an oblate spheroid (i.e., a disc with a=b and c<< a, b). The geometrical factors for a needle-shaped spheroid are $L_1 = 0$ and $L_2 = L_3 = \frac{1}{2}$, while those for a disc-shape spheroid are $L_1 = L_2 = 0$ and $L_3 = 1$. Once the polarizability is expressed in terms of $\varepsilon$, the CM relation is utilized to express the cluster polarizability in terms of the atomic polarizability. For example, for a sphere (a=b=c) in a vacuum, the polarizability is isotropic,

$$\alpha_{sphere} = a^3 \frac{\varepsilon - 1}{\varepsilon + 2} \qquad . \tag{15}$$

The CM relation [9, 14-16] for the static polarizability of a dielectric material, with number density $n_s \equiv \frac{N}{V}$, where $N$ and $V\left(= \frac{4\pi}{3} a^3 \right)$ are the total number of atoms and the volume of the dielectric sphere, is

$$\frac{\varepsilon - 1}{\varepsilon + 2} = \frac{4\pi}{3} n_s \alpha_{atom} = \frac{N}{a^3} \alpha_{atom} \qquad . \tag{16}$$

From eqs. (15) and (16), we obtain

$$\alpha_{sphere} = N \alpha_{atom} \qquad . \tag{17}$$

Hence, the enhancement factor $f$ of a dielectric sphere equals 1. The analytic expression for the static polarizabilities and the enhancement factors for other shapes of spheroidal dielectric are similarly obtained and listed in Table II.

The values of $f$ from the analytic expressions in Table II for two large-size limit cases are in good agreement with the asymptotic values for all shapes (see Table III). The agreement between the asymptotic values of the polarizability of a large cluster ($N \approx 1000$) and those determined from the analytic expressions for infinitely large clusters, indicates that a cluster larger than $N \approx 1000$ may be considered to be a continuum and the CM relation applies. For clusters of smaller size, however, the discrete nature of atoms comprising the cluster and the finite-size effect should be considered explicitly, as done in the present calculation, and the CM relation is not applicable. Note that there are larger discrepancies observed for sapphire clusters than other substances. This is because they have the largest value of $n_s \alpha_{atom}$ close to $3/(4\pi) \approx 0.24$ where the so-called "polarization catastrophe" occurs [9, 14-16]. At this point, the linear theory fails. Also, note that there is good agreement between results for a linear cluster (or a line) and the infinite chain model for all $n_s \alpha_{atom}$ (in both orientations). Between a square prism and the needle-shape dielectric, the agreement is excellent for perpendicular orientation and also for parallel orientation if $n_s \alpha_{atom}$ is small. The needle agrees better with the square prism than with the line, presumably because the former has a wider cross-section. Note that the continuum disc results for $f$ agrees very well with the numerical results for a square monolayer cluster (except for the large $n_s \alpha_{atom}$ case for parallel orientation). In contrast the results for the $L = \infty$ model of square monolayer disagree in all cases. This discrepancy is due to the neglect of the surface penetration region in the latter model calculations.

## III.    DISCUSSION AND CONCLUSION

Our study delineates the limits of the continuum and additivity approximations for polarizability.  In all cases studied, regardless of the shape and orientation of clusters, the continuum approximation held fairly well for $N$>1000, meaning that $f(N)$ reached an expected asymptotic value.  However, the additivity assumption ($f = 1$) is inconsistent with virtually all results. Additivity worked best for spheres and fairly well for the cubes. Even for sapphire (very high $n_s\alpha$), the asymptotic values of $f$ are close to one ($f = 1.018$ and $1.13$ for a sphere and a cube, respectively). Thus, $f = 1$ is a useful approximation for a sphere or a cube.

On the other hand, for an asymmetric cluster, such as a line or a square monolayer, the additivity approximation fails drastically. The enhancement factor is as high as 2 to 9 for a linear cluster parallel to an applied electric field (Figs. 2(a) and 3(a)). Interestingly, $f$ is not diminished by nearly this amount for linear clusters perpendicular to the applied electric field (Figs. 2(b) and 3(b)). That is, $f$ values do not go below 0.7, even for sapphire. This may be explained in relation to the arguments we made above with simple pairs of two identical dipoles. The dipole interaction in cases with $f>1$ has stronger magnitude (factor of 2) and thus extends effectively over larger range than that in cases with $f<1$, resulting in a larger |f-1|. Another interesting point noted is that when $f$ exceeds one, the function $f(N)$ has an inflection point near $N \lesssim 10$ in all cases. This inflection point is explained as the critical cluster size where the penetration depth becomes same order of magnitude with the cluster size.

One might wonder whether our approach to computing cluster polarizability is valid in an extreme case of a diatomic molecule. For example, one might naively try to approximate the polarizability of $H_2$ using the polarizability of an H atom ($\alpha_{atom} = 0.668$ Å$^3$) and the H-H bond length ($d = 0.74$ Å). However, putting these numbers into a two-atom calculation yields a negative cluster polarizability. Obviously, this result is not correct because it omits the strong chemical bond and its effect on the "atomic" polarizability. In order to examine the legitimacy of the dipolar approximation in other cases, we calculate the molecular polarizability for a number of diatomic molecules, based on the known atomic polarizability. The method proves to be accurate for low $n_s\alpha_{atom}$. First we assume that each molecule is made of two atoms (1 and 2) with separation distance $d$, and that the atoms interact such that

$$\mathbf{p}_1 = \alpha_1 \mathbf{E}_1 = \alpha_1 \left[ \mathbf{E}_0 + \frac{(3\mathbf{nn} - \mathbf{I}) \cdot \mathbf{p}_2}{d^3} \right] = \alpha_1 \mathbf{E}_0 + \gamma_1 (3\mathbf{nn} - \mathbf{I}) \cdot \mathbf{p}_2 \qquad (18)$$

$$\mathbf{p}_2 = \alpha_2 \mathbf{E}_2 = \alpha_2 \left[ \mathbf{E}_0 + \frac{(3\mathbf{nn} - \mathbf{I}) \cdot \mathbf{p}_1}{d^3} \right] = \alpha_2 \mathbf{E}_0 + \gamma_2 (3\mathbf{nn} - \mathbf{I}) \cdot \mathbf{p}_1 \qquad (19)$$

and

$$\gamma_i = \frac{\alpha_i}{d^3} \tag{20}$$

where $\mathbf{n}$ is the unit separation vector. As we noted before, when $n_s \alpha_{atom}$ is high getting close to the critical value of the polarization catastrophe, our method breaks down, and the molecule appears to be ferroelectric even when it is not. Substituting Eq. (19) into Eq. (18) and solving for $\mathbf{p}_1$ gives

$$\mathbf{p}_1 = [\mathbf{I} - \gamma_1 \gamma_2 (3\mathbf{nn} + \mathbf{I})]^{-1} \cdot [\alpha_1 \mathbf{I} + \gamma_1 \alpha_2 (3\mathbf{nn} - \mathbf{I})] \cdot \mathbf{E}_0 \quad . \tag{21}$$

Similarly,

$$\mathbf{p}_2 = [\mathbf{I} - \gamma_1 \gamma_2 (3\mathbf{nn} + \mathbf{I})]^{-1} \cdot [\alpha_2 \mathbf{I} + \gamma_2 \alpha_1 (3\mathbf{nn} - \mathbf{I})] \cdot \mathbf{E}_0 \quad . \tag{22}$$

From these, a molecular polarizability tensor for the cluster is obtained as

$$\boldsymbol{\alpha}_{cluster} = [\mathbf{I} - \gamma_1 \gamma_2 (3\mathbf{nn} + \mathbf{I})]^{-1} \cdot [(\alpha_1 + \alpha_1) \mathbf{I} + (\gamma_1 \alpha_2 + \gamma_2 \alpha_1)(3\mathbf{nn} - \mathbf{I})] \quad . \tag{23}$$

The molecule can rotate to all angles, and we will assume that all orientations are equally probable in an applied electric field. The final average polarizability of a diatomic molecule can then be represented as a scalar:

$$\langle \mathbf{p} \rangle = \langle p \rangle \mathbf{i}_z = \langle \mathbf{p}_1 + \mathbf{p}_2 \rangle = \langle \alpha_{cluster} \rangle \mathbf{E}_0 \tag{24}$$

The averaging integral is

$$\langle \alpha_{cluster} \rangle = \frac{\int\limits_0^{2\pi}\int\limits_0^{\pi} \boldsymbol{\alpha}_{cluster} \cdot \mathbf{i}_z \cdot \mathbf{i}_z \sin\theta \; d\theta \; d\phi}{4\pi} \tag{25}$$

Table IV lists the resulting values of $< \alpha_{cluster} >$ for some known molecular systems. We see that our model works fairly well for low $n_s \alpha_{atom}$. Those values for the model marked "---" are negative, indicating that our simple model fails in those molecules. Considering the simplicity of the model our method works surprising well for small $n_s \alpha_{atom}$.

There exist several applications of our results. One is the polarizability values themselves. A second is their use in computing VDW interactions, which we describe in detail elsewhere [18]. As an example of the nonadditivity effects on the VDW interactions, we describe one case, a spherical atom interacting with a long (L>100) linear cluster oriented parallel or perpendicular to their separation vector. At large separation, the interaction is said to be "fully retarded", in which case one needs to know only the static polarizability of the cluster. A factor $R$ describes the increase in the interactions compared to the result based on additivity, and

$$R = \frac{13 f_\perp + 10 f_{//}}{23} - 1 \ \text{ and } \ \frac{33 f_\perp + 13 f_{//}}{46} - 1 \ \text{ for parallel and perpendicular orientation,}$$

respectively. Using the result of the present calculation ($f_\perp \cong 0.78$ and $f_{//} \cong 2.2$), we obtain an enhancement of 40% and 20% for the respective orientations. This is a very interesting result, showing that the exact non-additivity effect is non-negligible and is always attractive, unlike the three-body interaction contribution to the non-additivity which is attractive and repulsive in parallel and perpendicular orientations, respectively. Thus, the three-body term may have a completely misleading sign.

One may also use the nonuniform distribution of polarization within a cluster obtained in the present calculation as an input to the discrete dipole approximation (DDA) method. The DDA was originally proposed by Purcell and Pennypecker [19] and has been extensively used to calculate scattering and absorption of electromagnetic waves by targets (or clusters) with arbitrary shapes [20]. In the DDA the electromagnetic properties of a scatterer are described by those of a collection of coupled electric dipoles residing on a cubic lattice sites. Therefore, the central quantity in the DDA is the electric dipole moments. Historically, in the DDA, a uniform polarization within a cluster has been assumed. However, there was a discrepancy between the spectra from DDA and those from experiments [20, 21]. Very recently, there has been an attempt to associate this discrepancy with the possible non-uniformity of the polarization within a cluster [21], which is exactly the result we obtained in the present study. In principle, one can use the microscopic method presented here to obtain the nonuniform distribution of polarization within an arbitrary shape cluster of interest and use these as input dipole moments for DDA-based applications.

In summary, our study reveals the limits of the continuum and additivity approximations for the static case of nonpolar nano-size systems. We also found that for a finite-size cluster, the surface effects due to the nonuniform local field near a surface play a very important role. We are currently studying more complex behavior, such as that for polar molecules (e.g., water). We are also evaluating the dynamic polarizability of finite-size clusters using a similar microscopic method, employing dynamical atomic polarizabilities. That study will allow us to examine all resonance peaks (including bulk and surface plasmons) in the cluster and shed light on the scattering and absorption spectra of finite-size clusters of arbitrary shape.

## IV.        ACKNOWLEDGMENT

This research is supported by the National Science Foundation through NSF NER Grant No. CTS-0403646 and Ben Franklin Technology Center of Excellence in Nanoparticulate Science and Engineering.  We are grateful to Craig Bohren, Jerry Mahan and Amand Lucas for helpful discussions.

**TABLE CAPTIONS**

**Table I**. Atomic data used in the present study [5].

**Table II**. Analytic expressions for the enhancement factors for infinite size discrete clusters and ellipsoidal continuum clusters. Here, $f_{chain}^{//E}(1\times1\times\infty)$ and $f_{chain}^{\perp E}(\infty\times1\times1)$ are for infinitely long chain of identical dipoles lying along and perpendicular to the applied external field, respectively. Similarly, $f_{square}^{//E}(1\times\infty\times\infty)$ and $f_{square}^{\perp E}(\infty\times\infty\times1)$ are for an infinite number of identical dipoles residing on a two dimensional simple cubic lattice in two orientations with respective to the applied external field. Also, $f_{cube}(\infty\times\infty\times\infty)$ is for an infinite number of identical dipoles residing on a three dimensional simple cubic lattice. $f_{needle}^{//E}$, $f_{needle}^{\perp E}$, $f_{disc}^{//E}$, and $f_{disc}^{\perp E}$ are evaluated for extremely anisotropic ellipsoids (needle for prolate spheroid and disc for oblate spheroid) in two orientations with respective to the applied external field. $f_{sphere}$ is for a sphere.

**Table III**. The asymptotic values of the enhancement factor $f$. (a) and (b) are for orientations of cluster with long dimension along and perpendicular to the applied field, respectively. All the notations are the same as in Table II with addition of $f_{line}^{//E}(1\times1\times1000)$, $f_{line}^{\perp E}(1000\times1\times1)$, $f_{prism}^{//E}(2\times2\times250)$, $f_{prism}^{\perp E}(250\times2\times2)$, $f_{square}^{//E}(1\times60\times60)$, and $f_{square}^{\perp E}(60\times60\times1)$ for finite size clusters (line, square prism and square monolayer, respectively) in two orientations with respect to the applied external field.

**Table IV.** Average diatomic molecular polarizability from experiment compared with our simple model. The results are listed in order of increasing polarizability density γ, defined in Eq.(20). All the parameters and the experimental polarizabilities are from Ref. 17.

**FIGURE CAPTIONS**

**Figure 1.** (Color online) Enhancement factor $f$ as a function of number of atoms $N=L^3$ for cubic clusters with dimension $L \times L \times L$. The symbols are open circle for sapphire, open triangle for silica and closed triangle for hexane.

**Figure 2.** (Color online) Enhancement factor as function of $N=L$ for linear clusters (with dimension $1 \times 1 \times L$). The symbols are open circle for sapphire, open triangle for silica and filled triangle for hexane. Case (a) corresponds to the long dimension of the cluster parallel to the applied field while case (b) refers to perpendicular orientation.

**Figure 3.** (Color online) Enhancement factor as function of $N=4L$ for square prism clusters (with dimension $2 \times 2 \times L$). The symbols are open circle for sapphire, open triangle for silica and filled triangle for hexane. Case (a) corresponds to long dimension of the cluster parallel to the applied field while case (b) refers to perpendicular orientation.

**Figure 4.** (Color online) Enhancement factor as function of $N=L^2$ for square monolayer clusters ($1 \times L \times L$).The symbols are open circle for sapphire, open triangle for silica and filled triangle for hexane. Case (a) corresponds to the long dimension of the cluster parallel to the applied field while case (b) refers to perpendicular orientation.

**Figure 5.** (Color online) Magnitude distribution of the induced dipole moments in a square silica monolayer (60x60x1) when it lies perpendicular to the external field as in Fig. 4(b). Here, the external field is pointing vertically into the paper. Darker (blue in color) region is for larger magnitude. Refer to Fig. 8 for numerical values.

**Figure 6.** (Color online) Magnitude distribution of the induced dipole moments in a square silica monolayer (60x1x60) when it lies along the external field as in Fig. 4(a). The external field is pointing horizontally into the right in this figure. Darker (blue in color) region is for larger magnitude. Refer to Fig. 9 for numerical values.

**Figure 7.** (Color online) Orientation distribution of induced dipole moments in a square silica monolayer (60x1x60) when it lies along the external field as in Fig. 4(a). The external field is pointing horizontally into the right (z-axis) in this figure. The angle is defined as $\tan^{-1}(p_x/p_z)$, where $p_x$ and $p_z$ are the x (vertical) and z (horizontal) component of the induced dipole moment. (a) The angles are positive in the upper right and the lower left corners and negative in the upper left and the lower right corners. Darker region in the corners is for larger magnitude of angle. (b) The relative orientation and magnitude of dipoles are schematically drawn with arrows.

**Figure 8.** (Color online) The local enhancement factor as a function of the location along the center line on square silica monolayers of various sizes (LxLx1). Here, the square monolayer lies perpendicular to the electric field. L=5 (dotted line with ◆), 10 (dashes with ■), 20 (◇), 30 (●), and 60 (✶).

**Figure 9.** (Color online) The local enhancement factor as function of the location along the center line on square silica monolayers of various sizes (Lx1xL); (a) the center line is perpendicular to the external field and (b) the center line is parallel to the field. Here, the square monolayer lies along the electric field and same symbols are used for each L as in Fig. 8.

**Figure 10.** (Color online) The orientation of the induced dipole moments as function of the location along the top horizontal edge on square silica monolayers of various sizes (Lx1xL). The orientation angle is defined in the figure caption of Fig. 7. Here, the square monolayer lies along the electric field and same symbols are used for each L as in Fig. 8.

**Figure 11.** (Color online) The local enhancement factor as function of the location along the linear silica clusters of various lengths (Lx1x1). Here, the linear cluster lies perpendicular to the electric field. L=2 (closed square with broken line), 3 (open diamond with dotted line), 4 (closed diamond with dotted line), 5 (open triangle with solid line), 10 (closed triangle with solid line), 50 (open diamond with solid line), 100 (cross), 400 (dotted line with no symbol), and 1000 (solid line with no symbol).

**Figure 12.** (Color online) The local enhancement factor as function of the position along the linear silica clusters of various lengths (1x1xL). Here, the linear cluster lies parallel to the electric field. (a) Same symbols are used for each L as in Fig. 11; (b) The x-axis is the distance from the left hand side edge into the cluster (in units of $a_0$) for L=50. The local enhancement factors for L>50 coincide with those of L=50 within the resolution of this figure.

**Figure 13.** Enhancement factor of a spherical cluster as a function of radius(a) and as a function of the inverse of the radius(b). Here, the radius R* is in unit of 5.665Å. Dotted lines are drawn to guide the eye. The symbols are open circle for sapphire, open triangle for silica and filled triangle for hexane.

**Table I.  H.-Y. Kim et.al.**

| substance | $\varepsilon$ | $a_0\,(\overset{\circ}{A})$ | $n_s\,(\overset{\circ}{A}^{-3})$ | $\alpha_{atom}\,(\overset{\circ}{A}^3)$ | $n_s\alpha_{atom}$ |
|-----------|------|-------|---------|-------|-------|
| hexane | 1.89 | 6.009 | 0.00461 | 11.85 | 0.055 |
| silica | 3.81 | 3.569 | 0.0220 | 5.25 | 0.115 |
| sapphire | 11.58 | 3.486 | 0.0236 | 7.88 | 0.186 |

**Table II.  H.-Y. Kim et.al.**

| infinite size cluster | ellipsoidal continuum cluster |
|---|---|
| $f_{chain}^{//E}(1 \times 1 \times \infty) = \left[1 - 4\zeta(3)n_s\alpha_{atom}\right]^{-1}$ | $f_{needle}^{//E} = \left[1 - \left(\dfrac{4\pi}{3}\right)n_s\alpha_{atom}\right]^{-1}$ |
| $f_{chain}^{\perp E}(\infty \times 1 \times 1) = \left[1 + 2\zeta(3)n_s\alpha_{atom}\right]^{-1}$ | $f_{needle}^{\perp E} = \left[1 + \left(\dfrac{2\pi}{3}\right)n_s\alpha_{atom}\right]^{-1}$ |
| $f_{square}^{//E}(1 \times \infty \times \infty) = \left[1 - \left(\dfrac{4\pi^2}{9}\right)n_s\alpha_{atom}\right]^{-1}$ | $f_{disc}^{//E} = \left[1 - \left(\dfrac{4\pi}{3}\right)n_s\alpha_{atom}\right]^{-1}$ |
| $f_{square}^{\perp E}(\infty \times \infty \times 1) = \left[1 + \left(\dfrac{2\pi^2}{3}\right)n_s\alpha_{atom}\right]^{-1}$ | $f_{disc}^{\perp E} = \left[1 + \left(\dfrac{8\pi}{3}\right)n_s\alpha_{atom}\right]^{-1}$ |
| $f_{cube}(\infty \times \infty \times \infty) = 1$ | $f_{sphere} = 1$ |

## Table III(a). H.-Y. Kim et. al.

| substance | $f_{line}^{//E}$ $(1 \times 1 \times 1000)$ | $f_{chain}^{//E}$ $(1 \times 1 \times \infty)$ | $f_{prism}^{//E}$ $(2 \times 2 \times 250)$ | $f_{needle}^{//E}$ | $f_{square}^{//E}$ $(1 \times 60 \times 60)$ | $f_{square}^{//E}$ $(1 \times \infty \times \infty)$ | $f_{disc}^{//E}$ |
|---|---|---|---|---|---|---|---|
| hexane | 1.35 | 1.36 | 1.32 | 1.30 | 1.30 | 1.21 | 1.30 |
| silica | 2.24 | 2.25 | 2.06 | 1.94 | 1.96 | 1.57 | 1.94 |
| sapphire | 9.33 | 9.45 | 5.97 | 4.52 | 4.92 | 2.41 | 4.52 |

## Table III(b). H.-Y. Kim et. al.

| substance | $f_{line}^{\perp E}$ $(1000 \times 1 \times 1)$ | $f_{chain}^{\perp E}$ $(\infty \times 1 \times 1)$ | $f_{prism}^{\perp E}$ $(250 \times 2 \times 2)$ | $f_{needle}^{\perp E}$ | $f_{square}^{\perp E}$ $(60 \times 60 \times 1)$ | $f_{square}^{\perp E}$ $(\infty \times \infty \times 1)$ | $f_{disc}^{\perp E}$ |
|---|---|---|---|---|---|---|---|
| hexane | 0.88 | 0.88 | 0.89 | 0.90 | 0.69 | 0.74 | 0.69 |
| silica | 0.78 | 0.78 | 0.80 | 0.81 | 0.51 | 0.58 | 0.51 |
| sapphire | 0.69 | 0.69 | 0.72 | 0.72 | 0.39 | 0.46 | 0.39 |

**Table IV. H.-Y. Kim et. al.**

| molecule | $\alpha_1 (\overset{\circ}{A}{}^3)$ | $\alpha_2 (\overset{\circ}{A}{}^3)$ | $d (\overset{\circ}{A})$ | $\gamma_1$ | $\gamma_2$ | $\langle\alpha\rangle_{expt}$ | $\langle\alpha\rangle_{model}$ |
|---|---|---|---|---|---|---|---|
| I$_2$ | 4.0 | 4.0 | 2.67 | 0.21 | 0.21 | 9.31 | 9.01 |
| Br$_2$ | 3.1 | 3.1 | 2.28 | 0.261 | 0.261 | 6.5 | 7.61 |
| Cl$_2$ | 2.1 | 2.1 | 1.99 | 0.266 | 0.266 | 4.6 | 5.21 |
| O$_2$ | 0.793 | 0.793 | 1.21 | 0.448 | 0.448 | 1.59 | 5.778 |
|  |  |  |  |  |  |  |  |
| HCl | 0.668 | 2.1 | 1.27 | 0.326 | 1.025 | 2.6 | --- |
| Na$_2$ | 24 | 24 | 3.08 | 0.821 | 0.821 | 30 | --- |
| N$_2$ | 1.10 | 1.10 | 1.10 | 1.00 | 1.00 | 1.77 | --- |
| CO | 1.75 | 0.793 | 1.13 | 1.213 | 0.55 | 1.95 | --- |
| H$_2$ | 0.668 | 0.668 | 0.74 | 1.65 | 1.65 | 0.803 | --- |

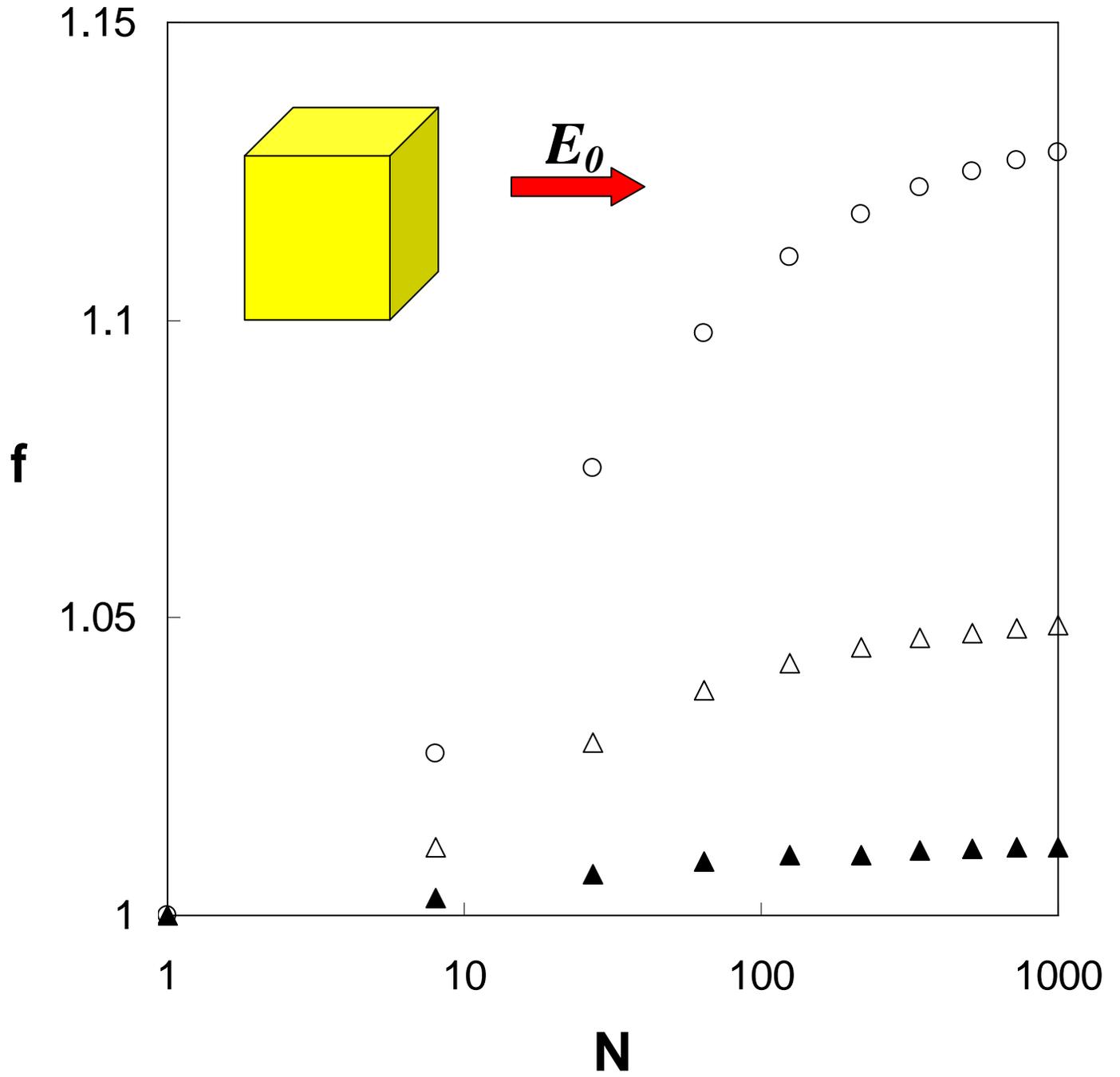

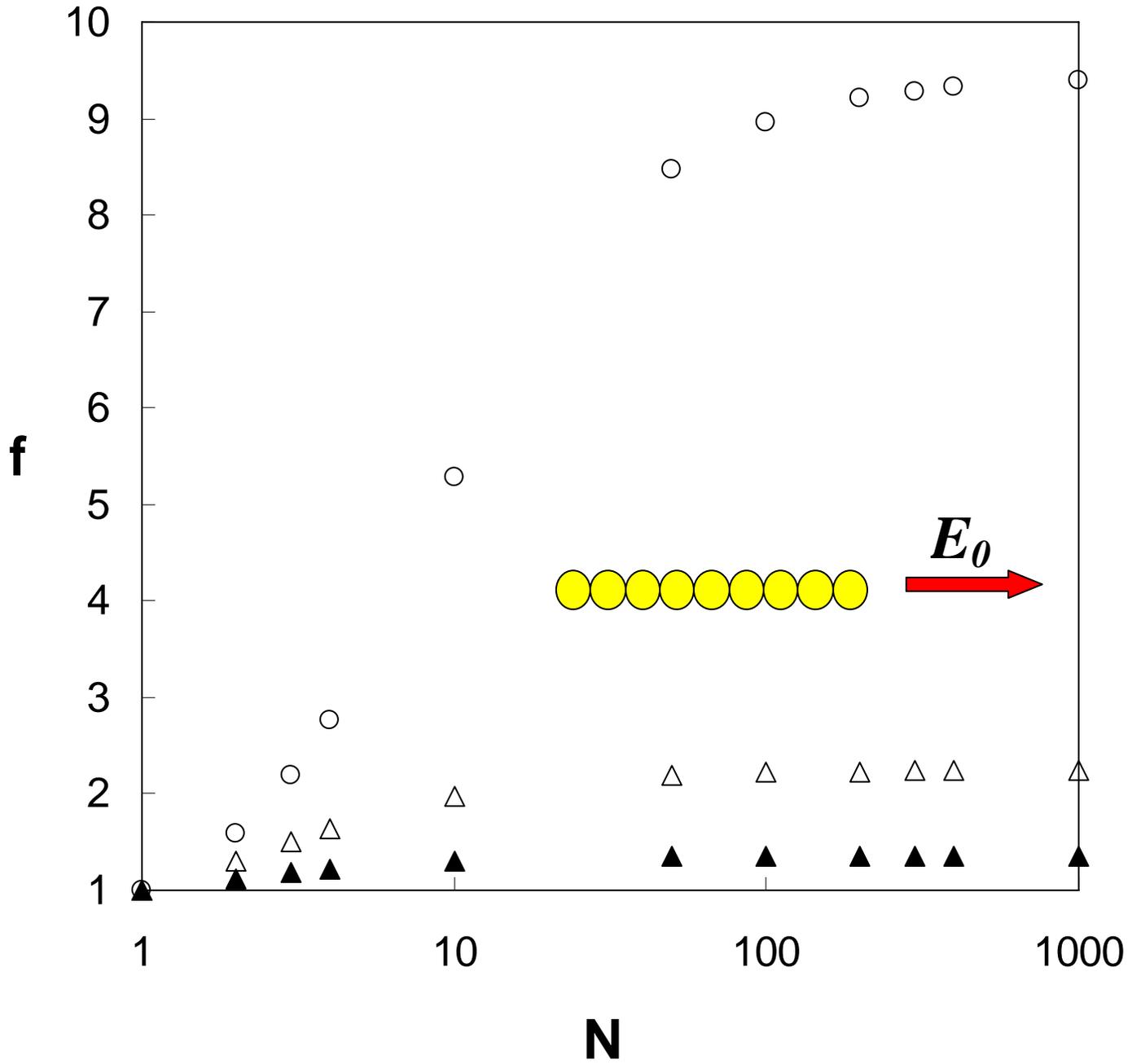

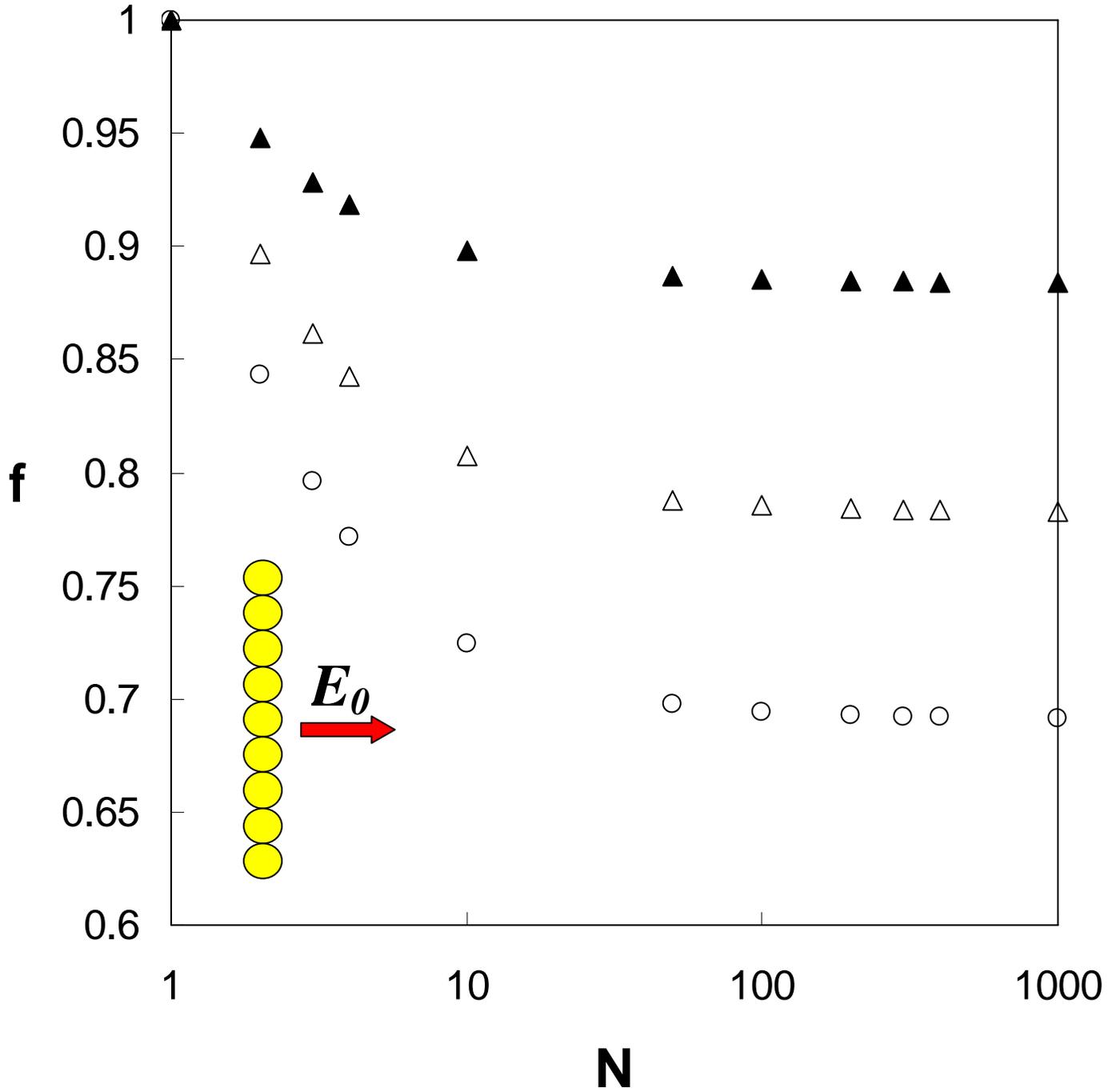

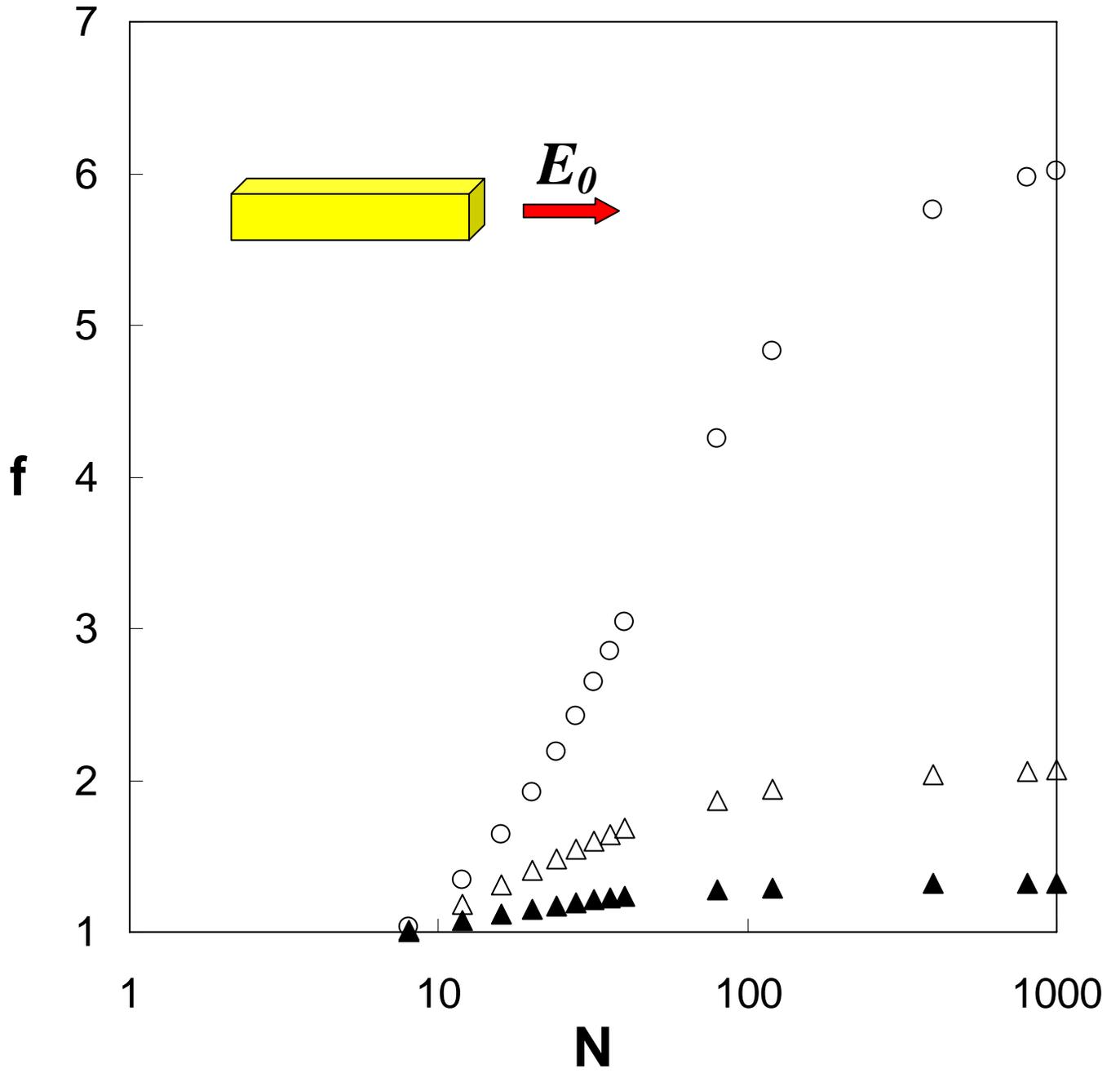

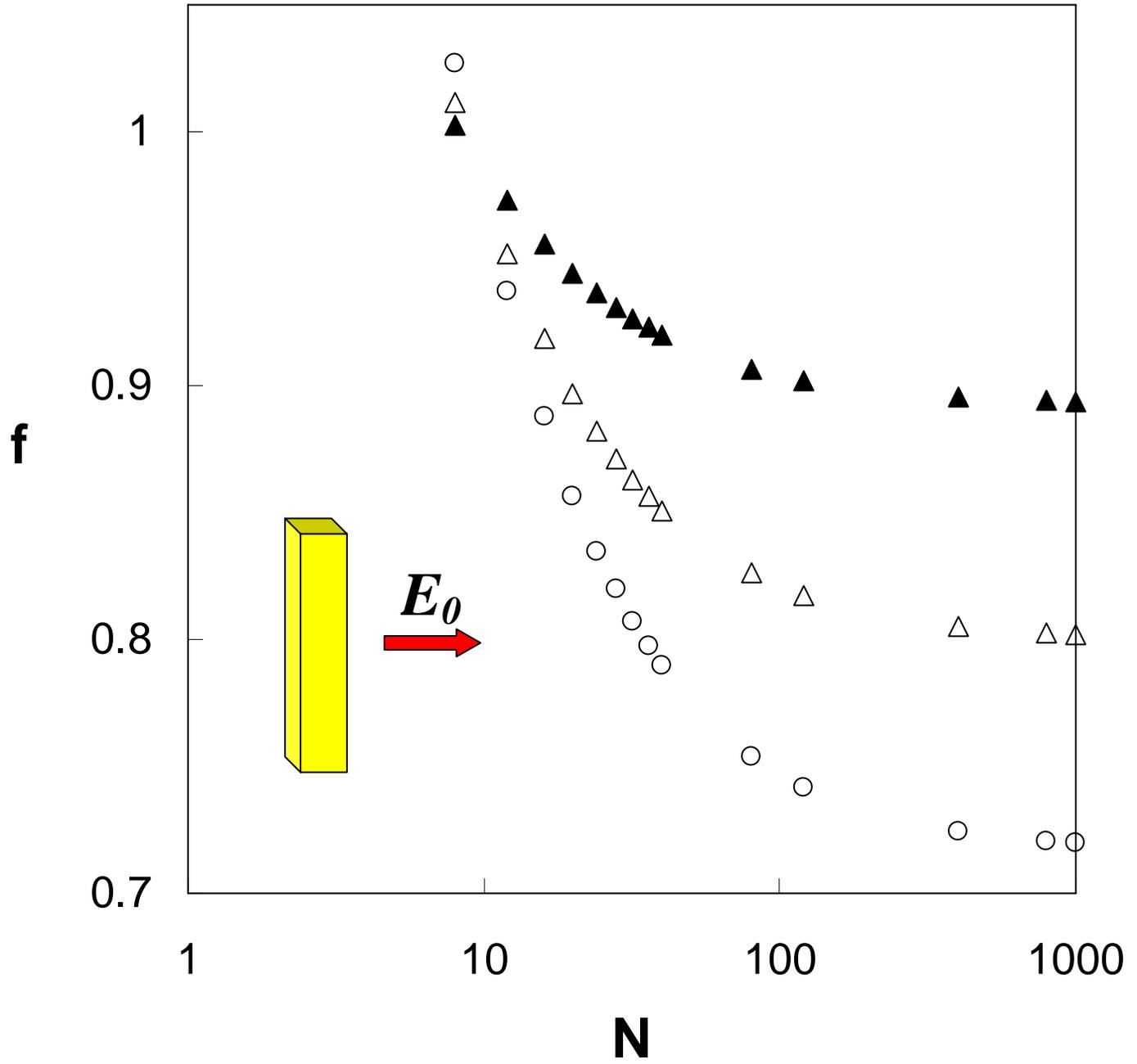

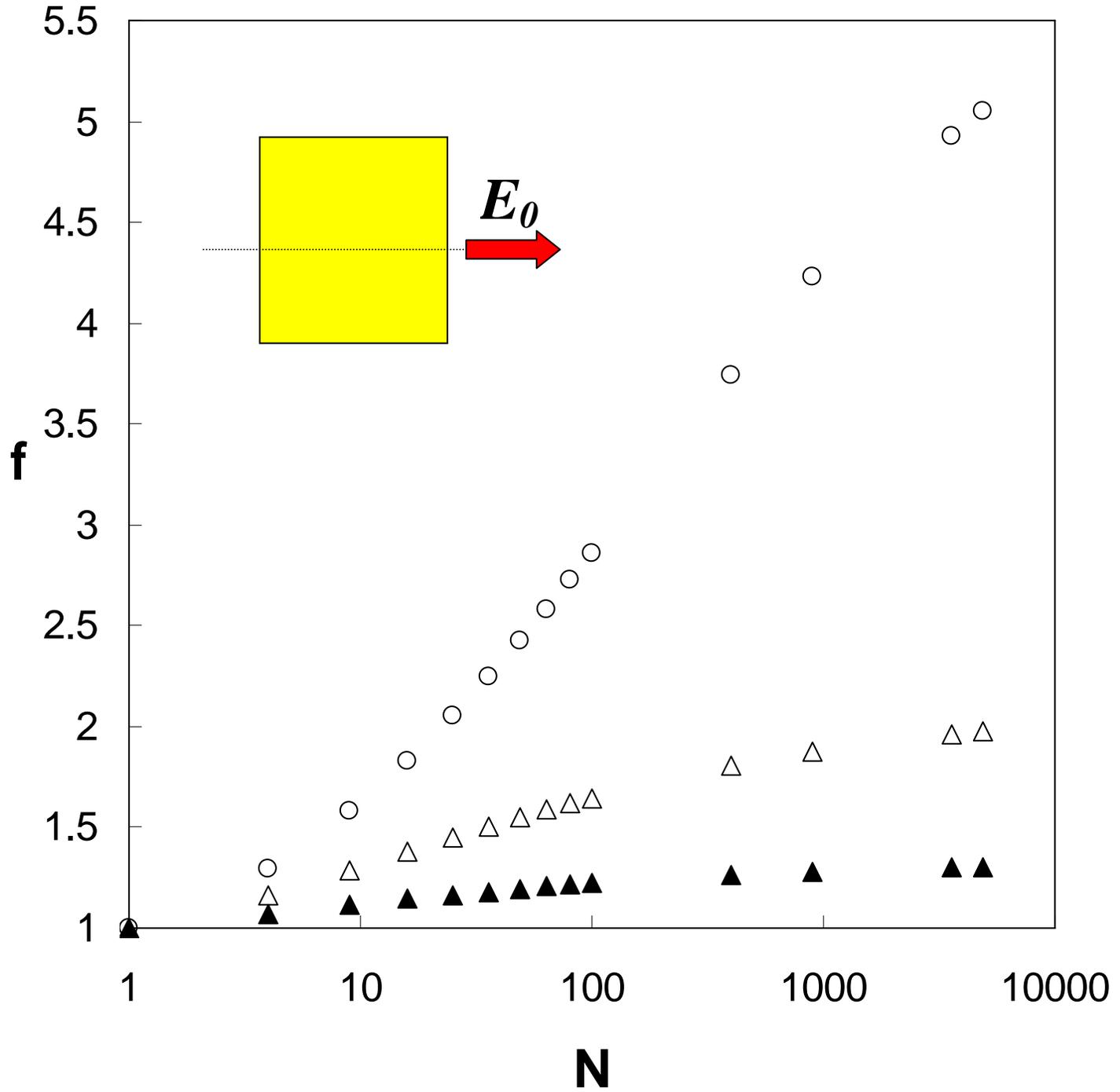

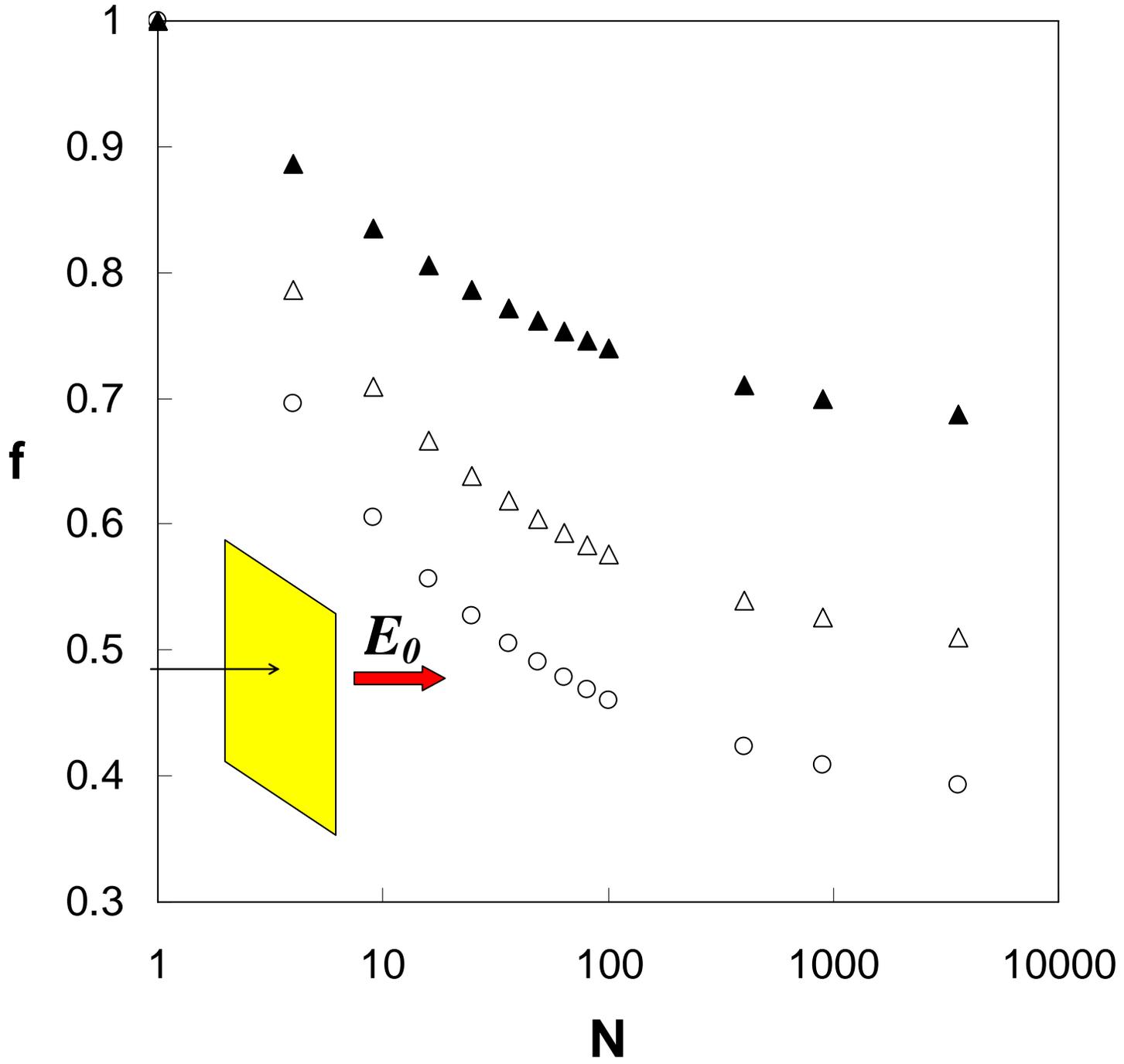

**FIGURE 5. H.-Y. Kim et. al**

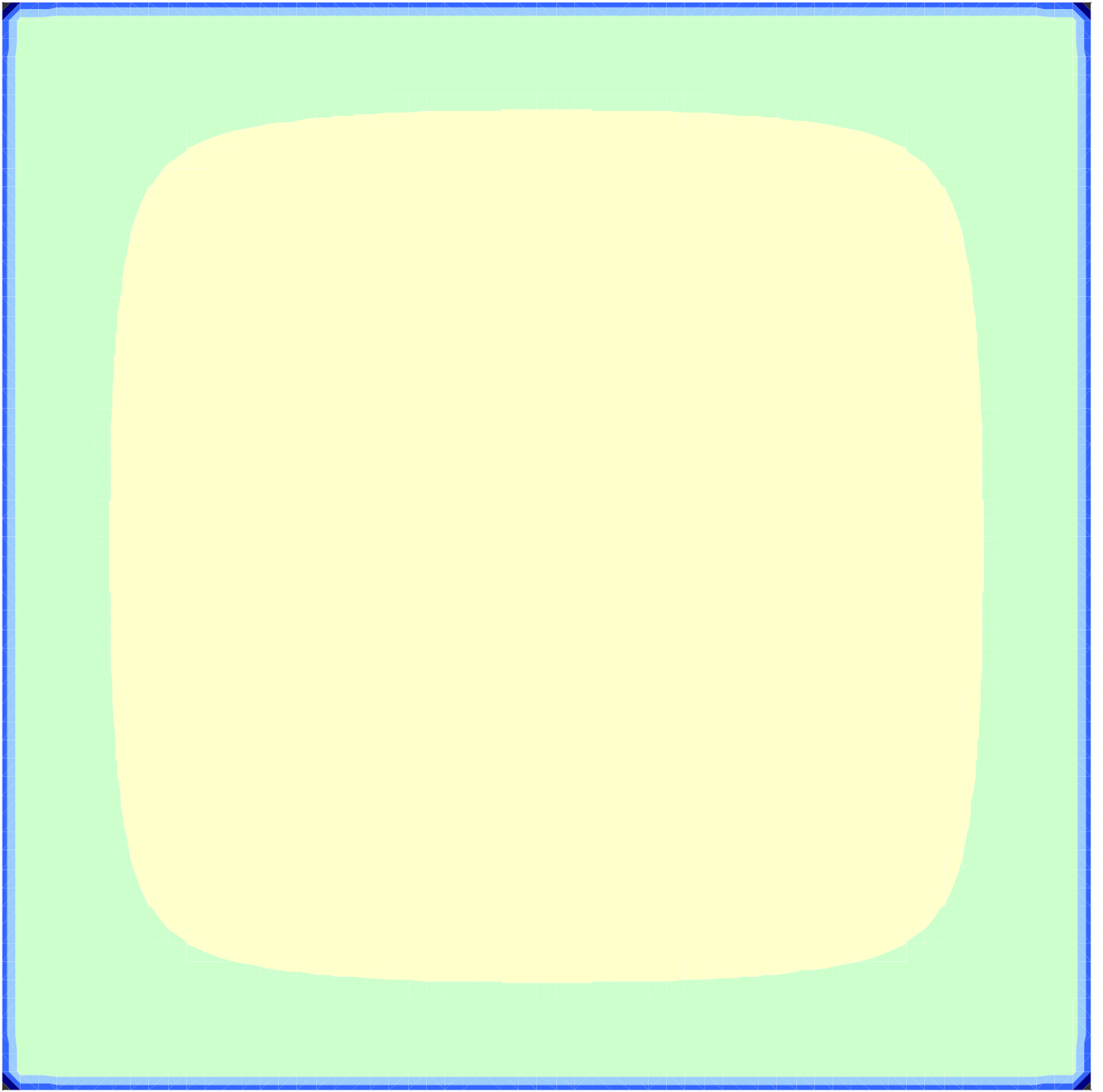

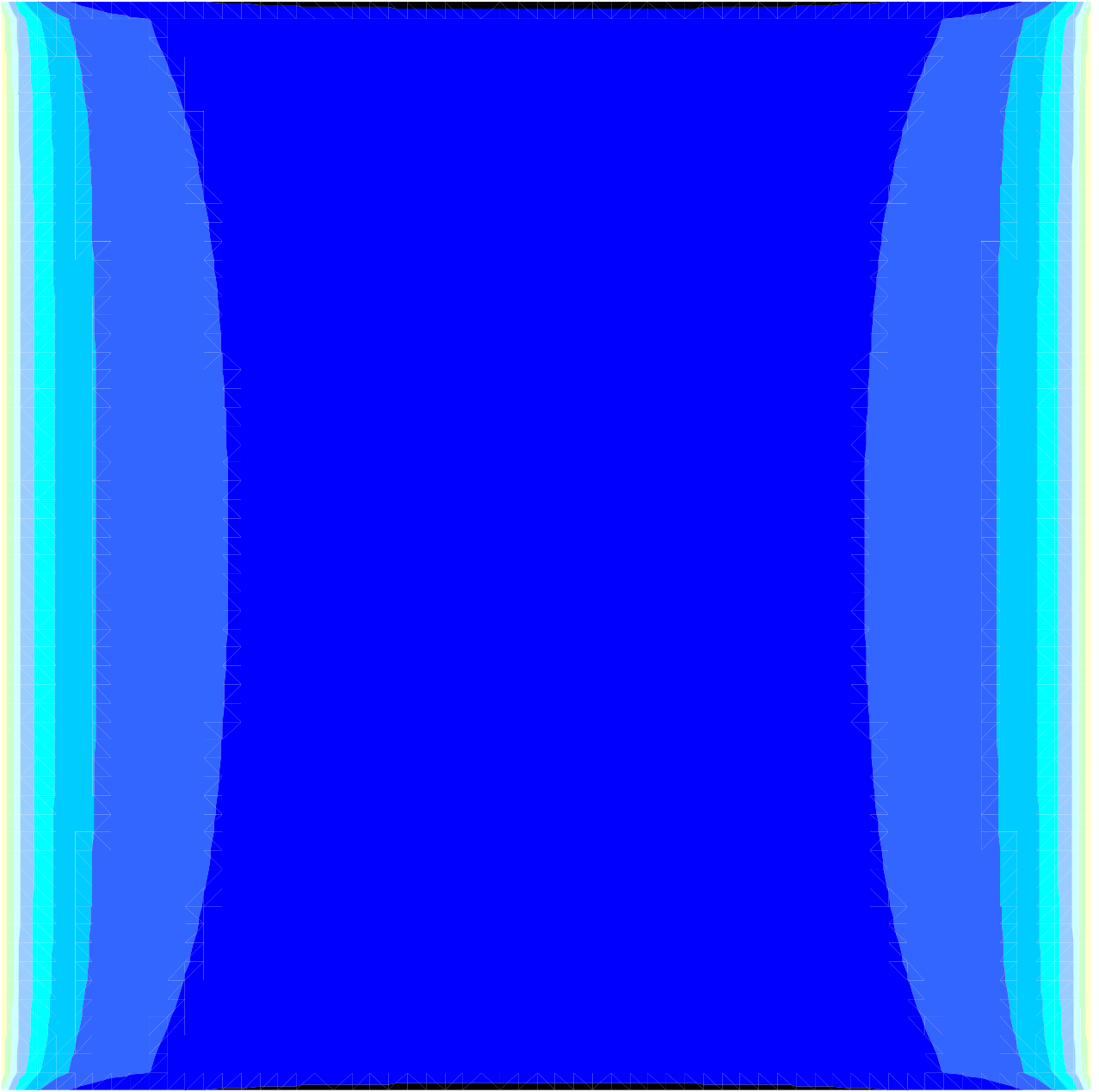

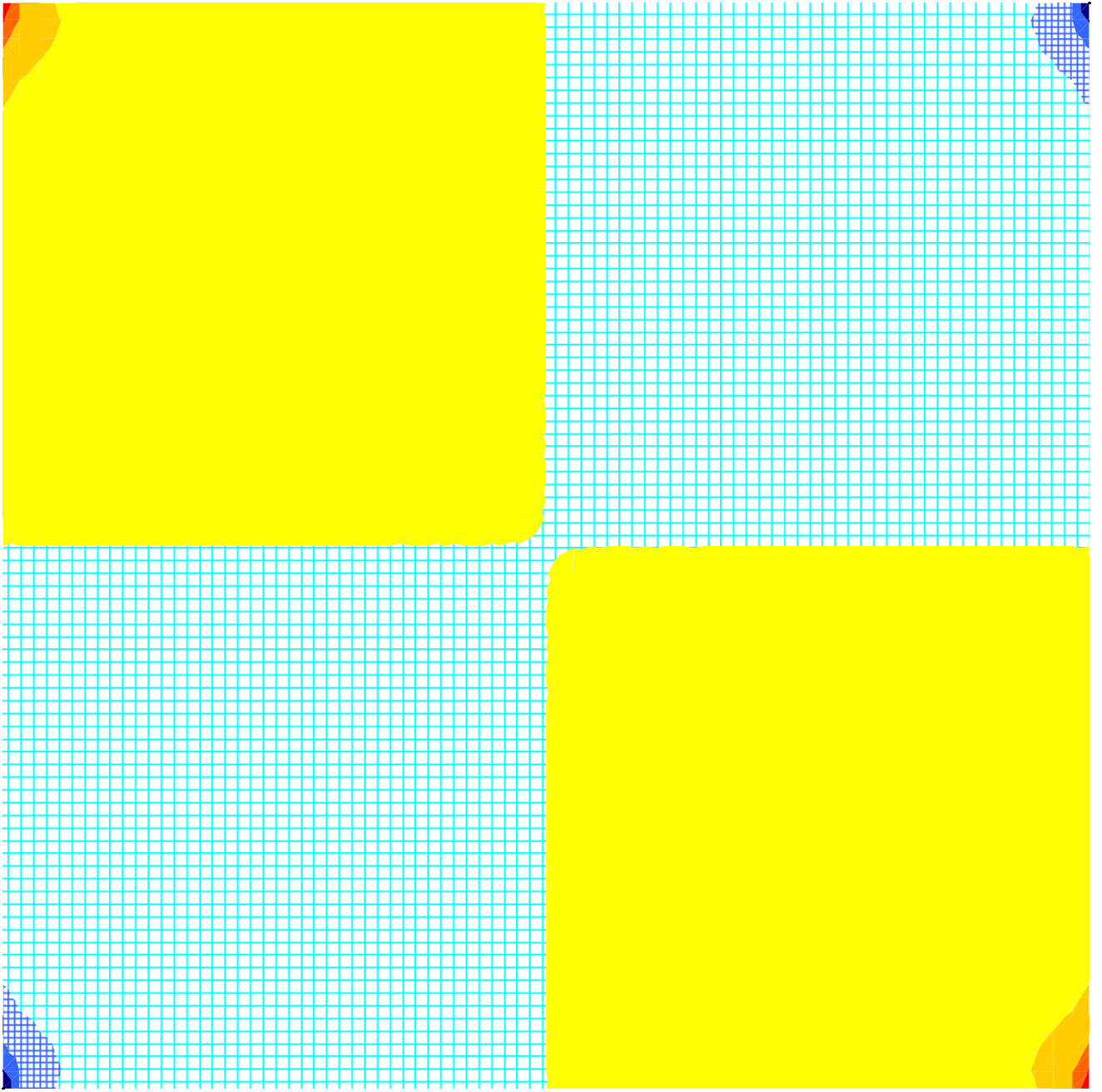

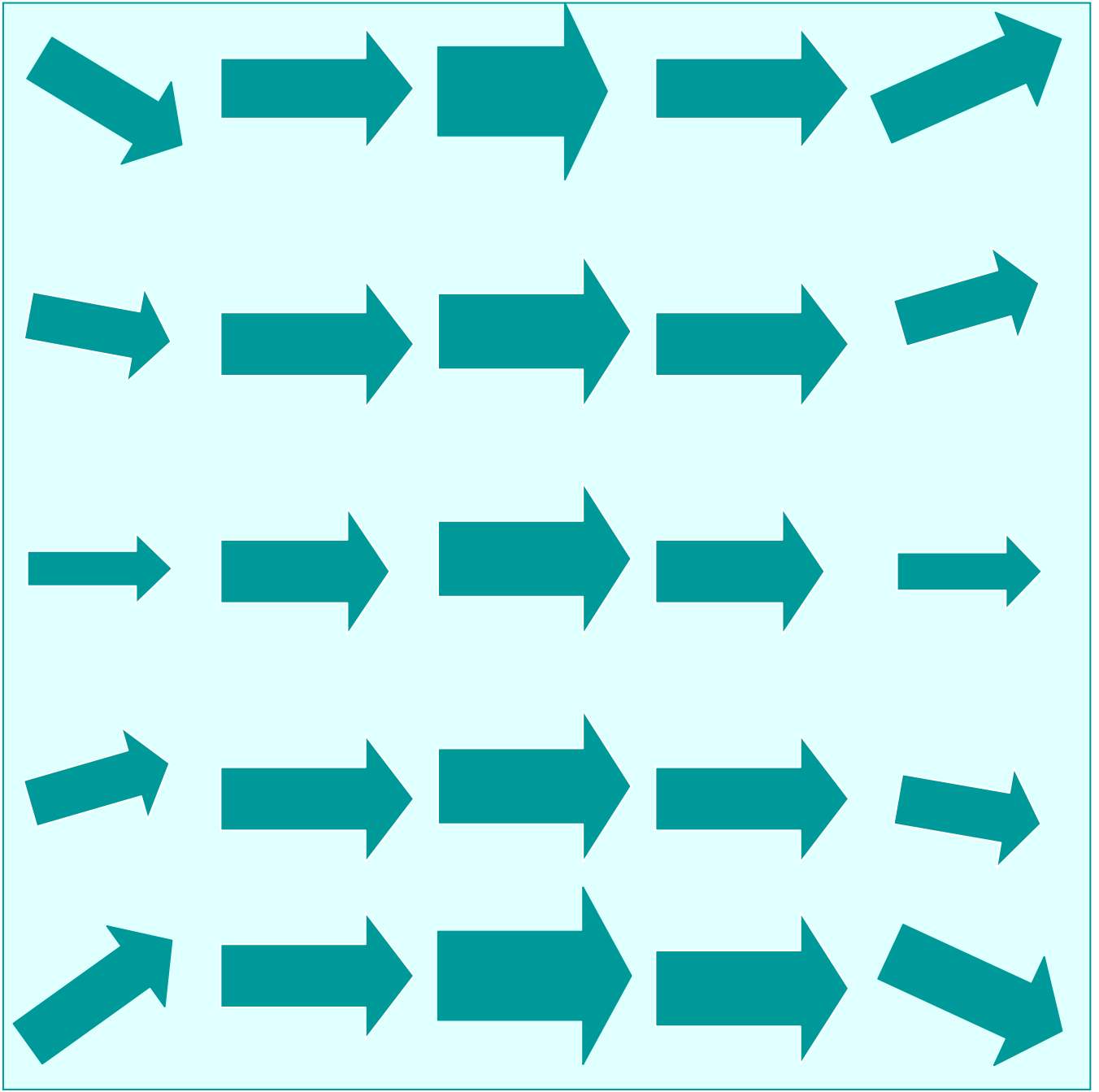

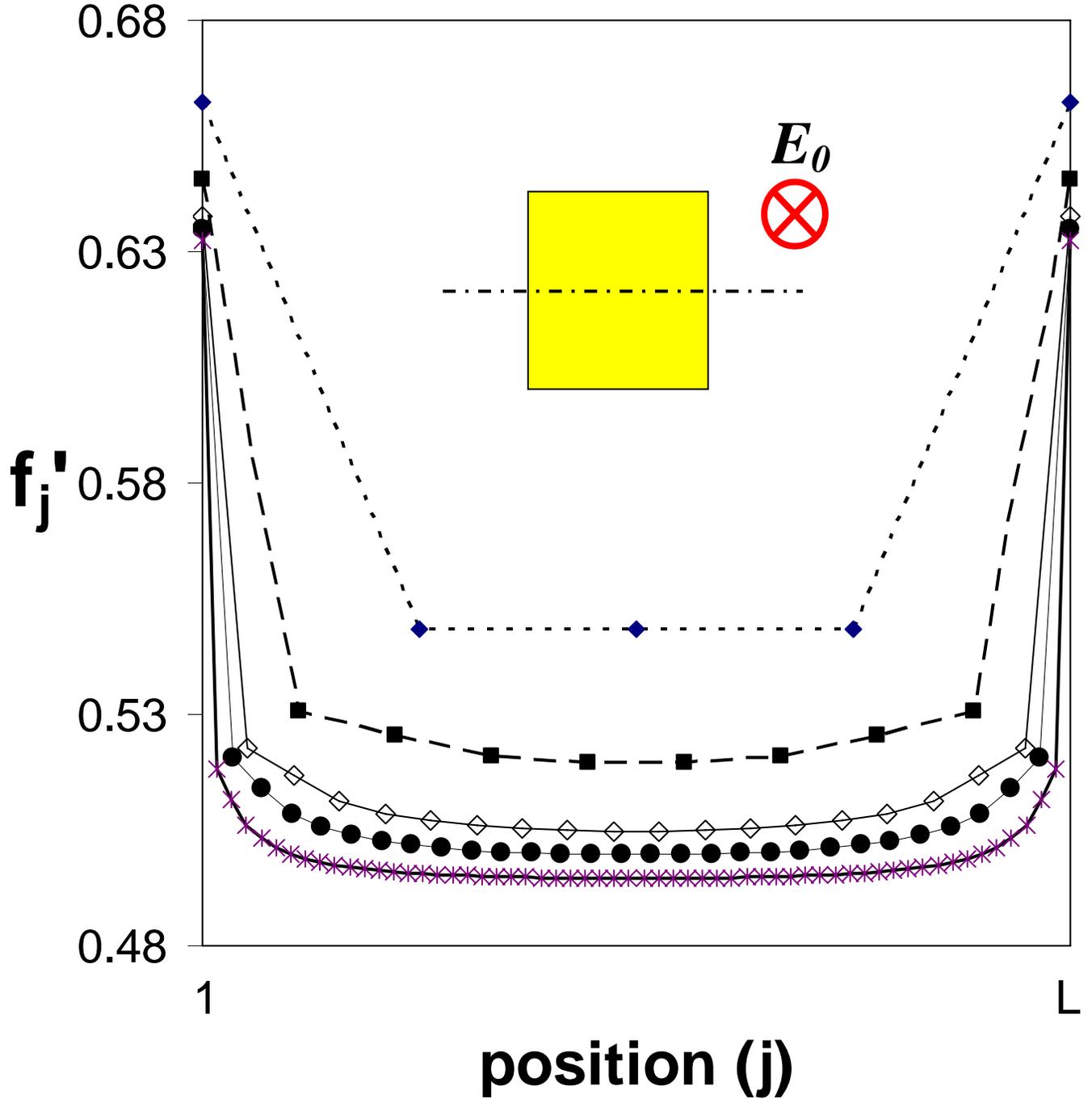

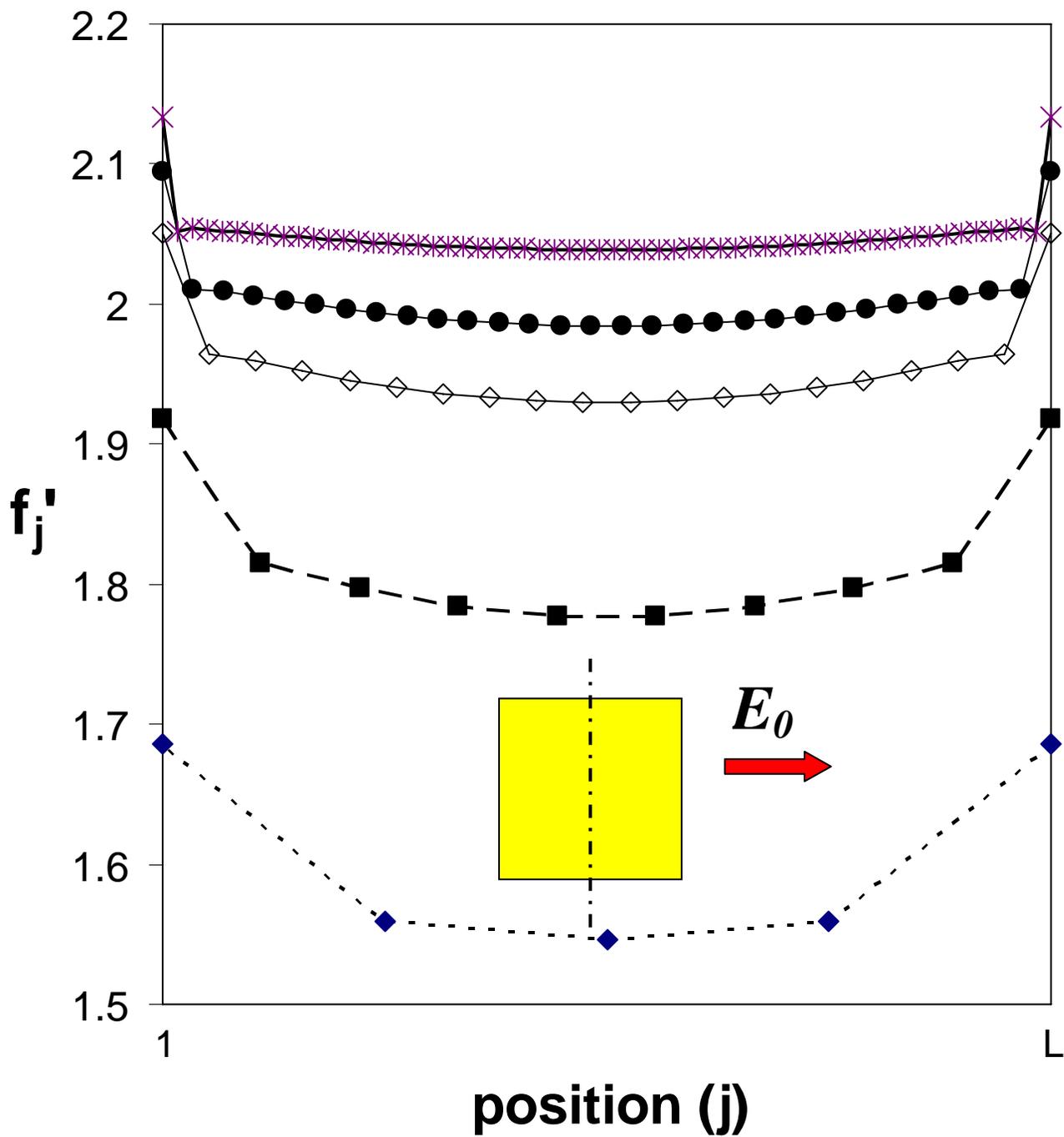

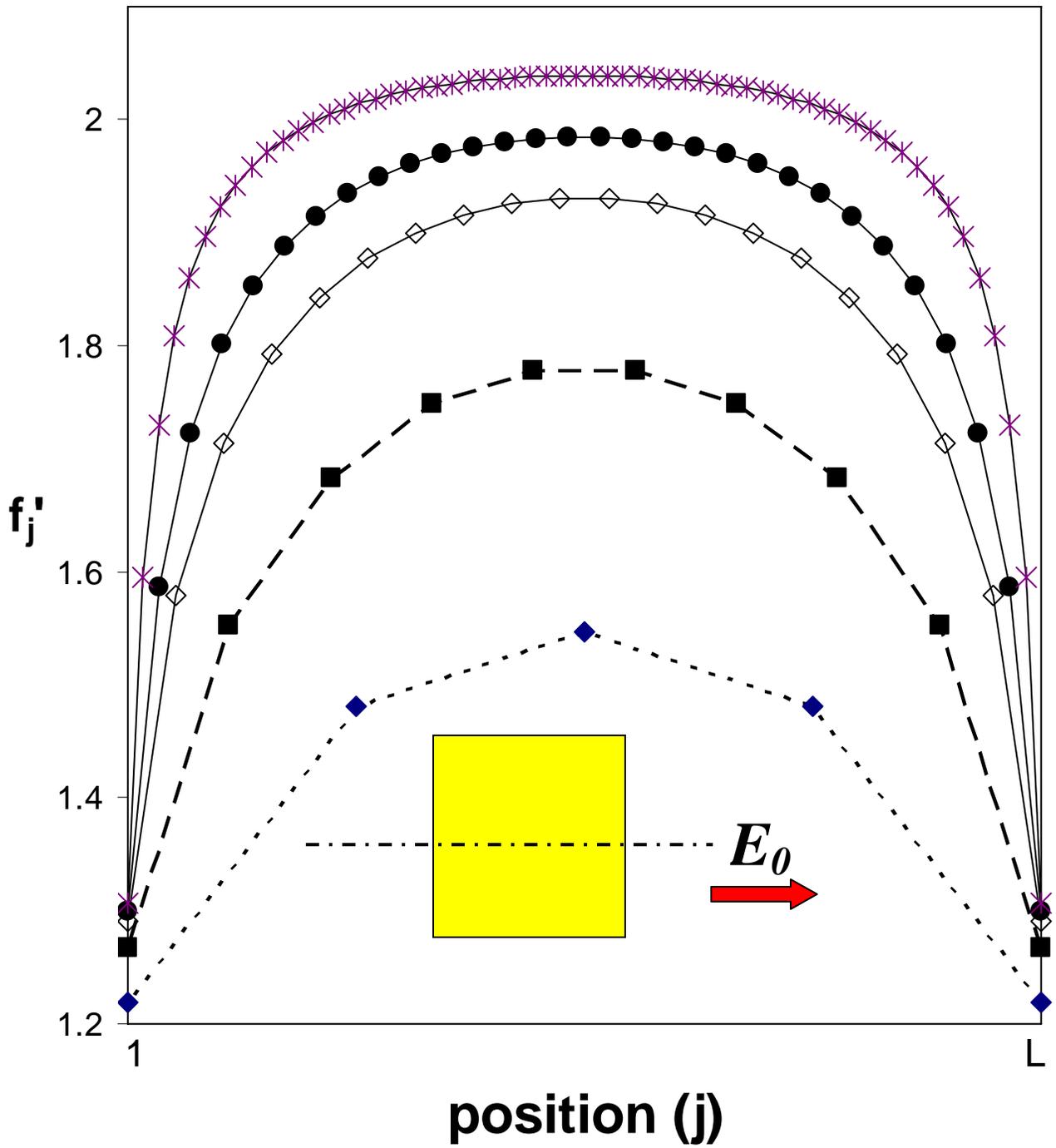

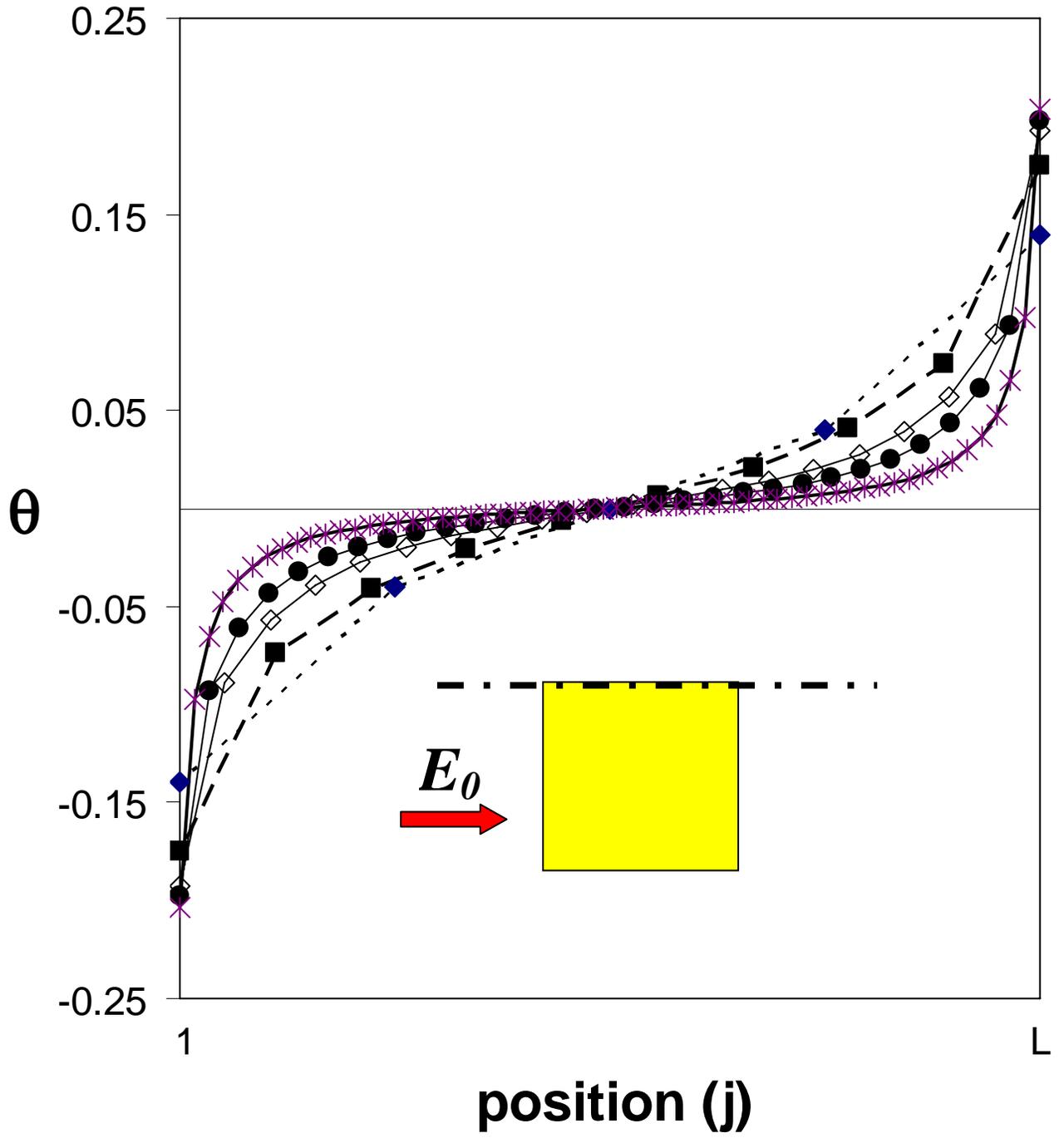

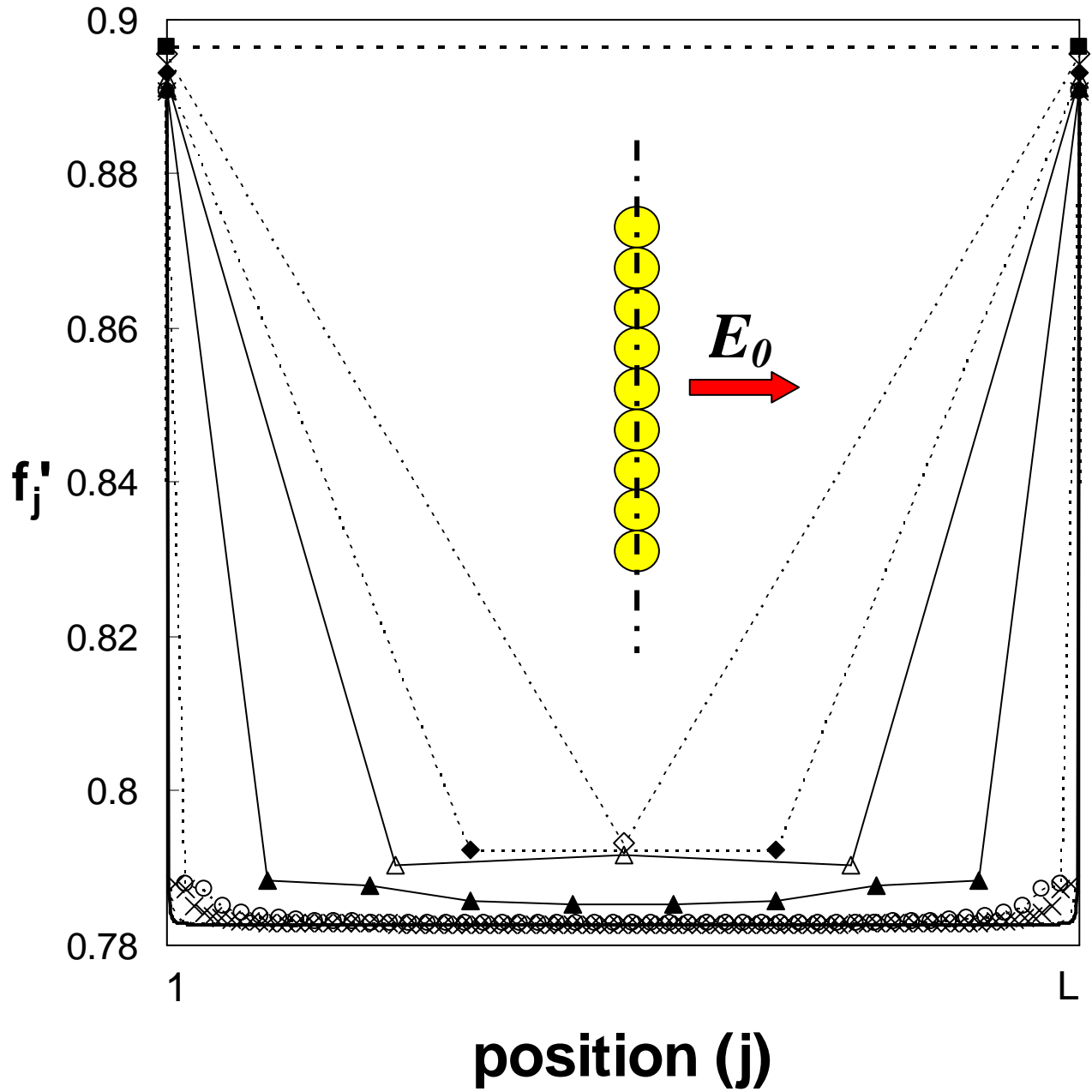

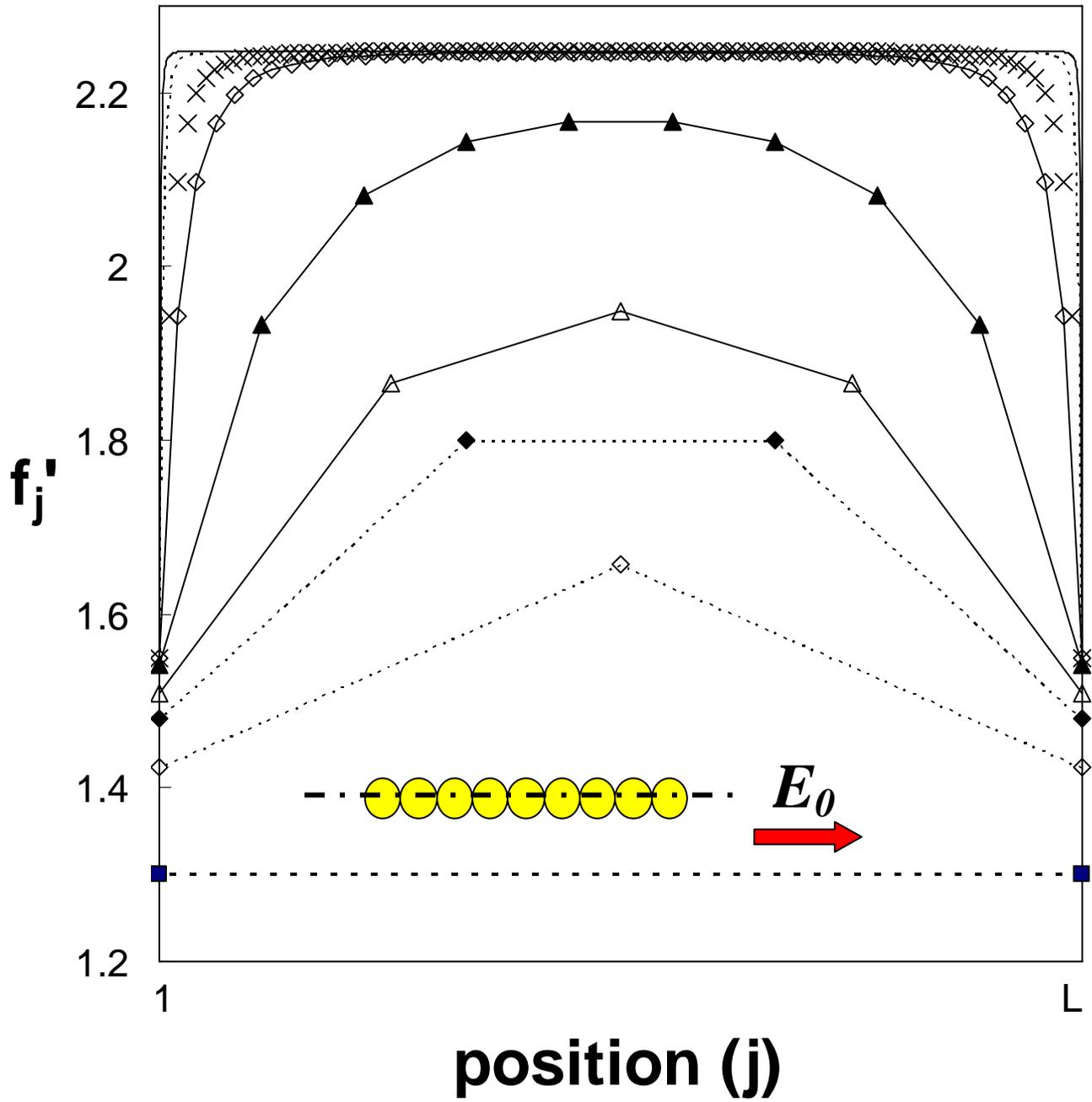

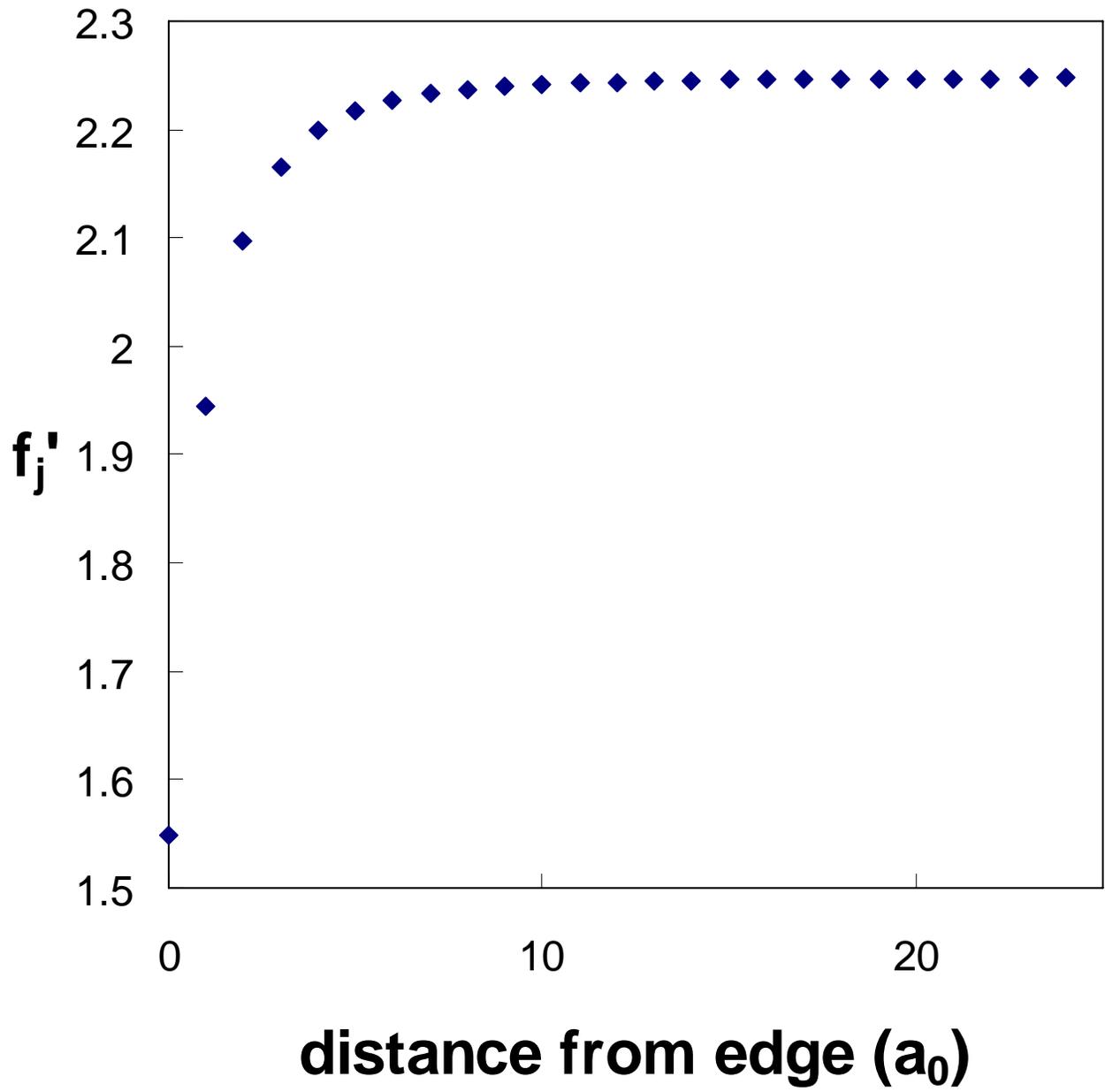

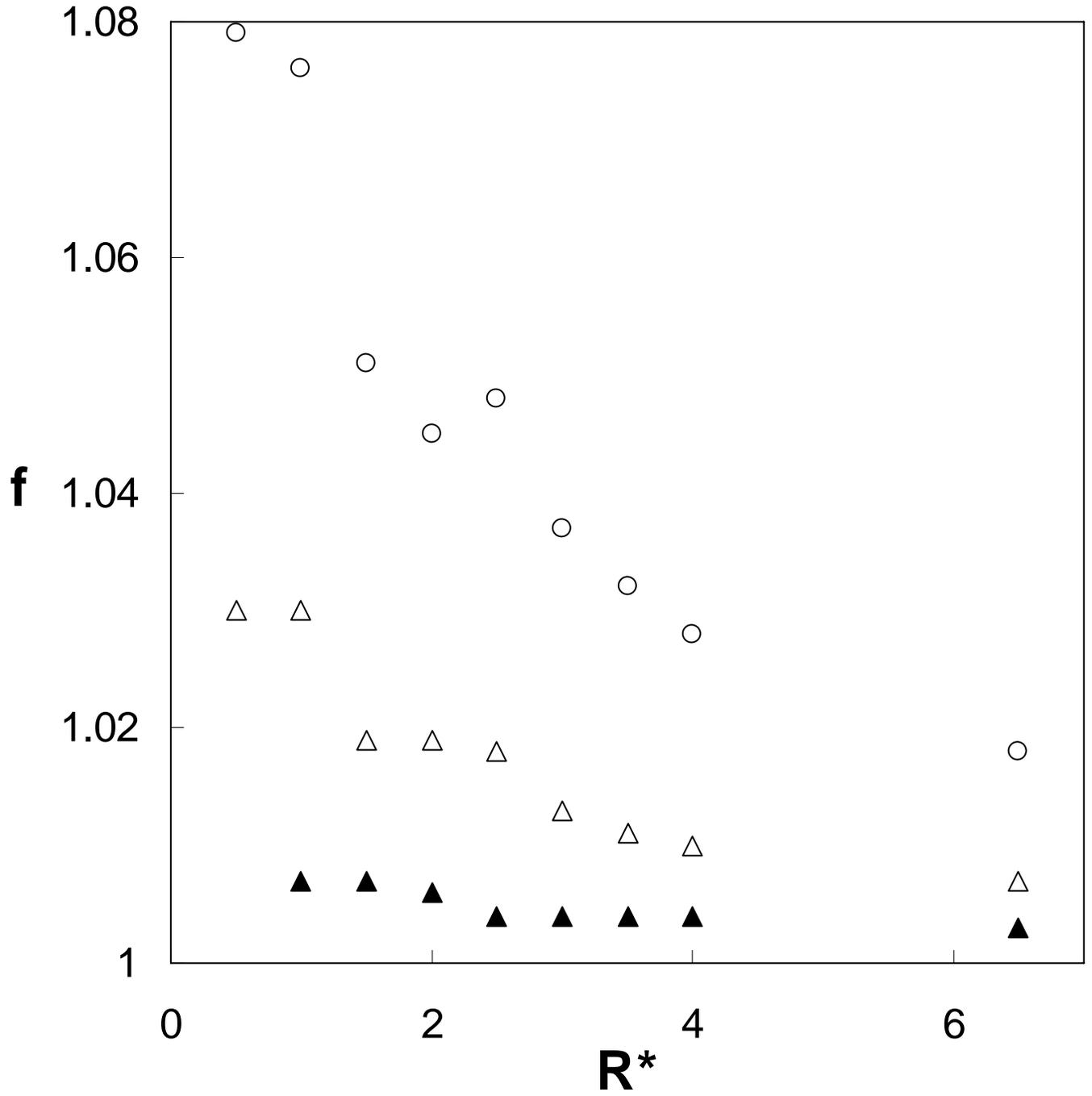

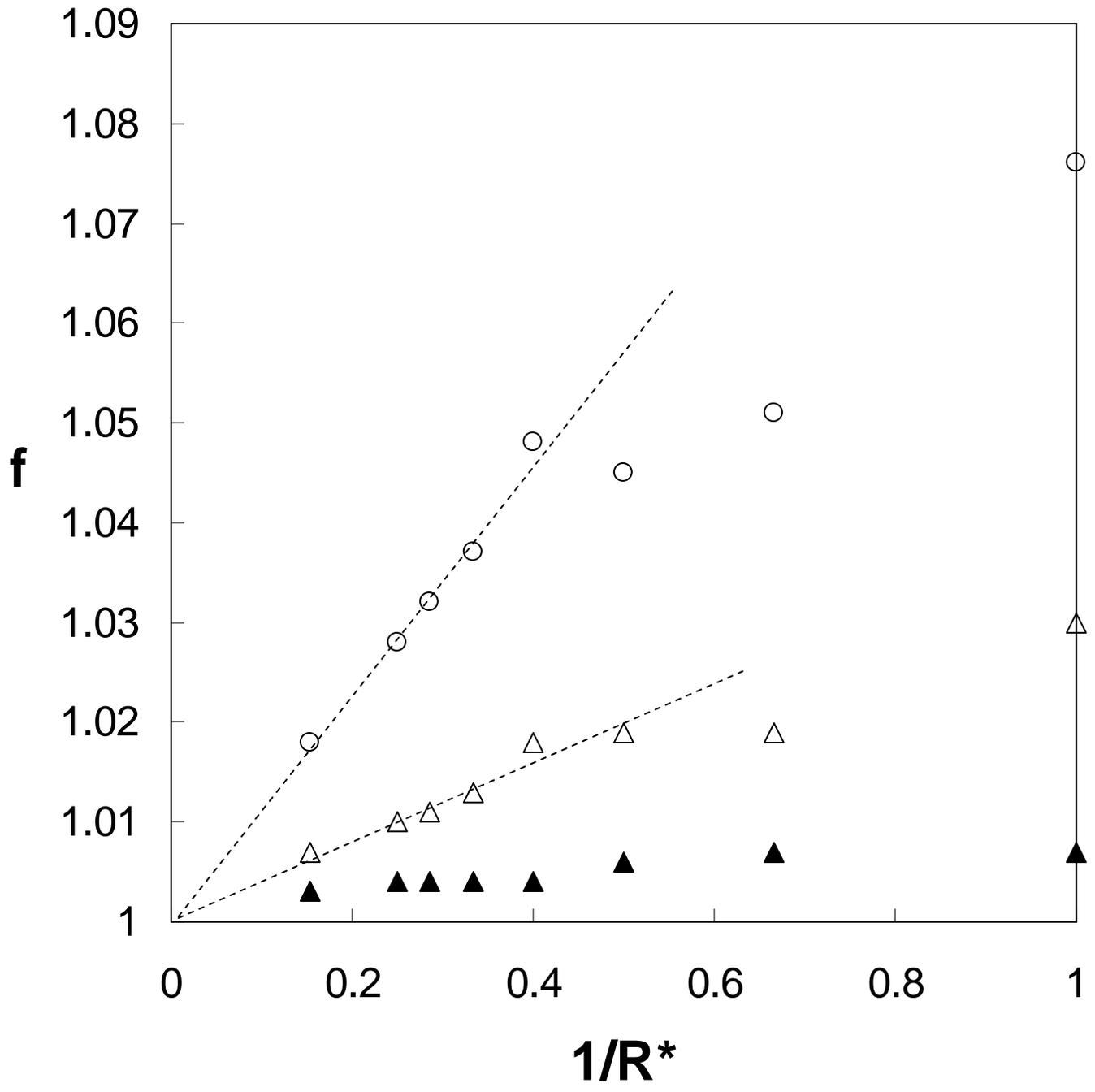